\documentclass[12pt,a4paper]{article}

\usepackage{amssymb}

\usepackage{graphics}

\makeatletter

\@addtoreset{equation}{section} \makeatother \newcommand{\anote}[1]{}
\newcommand{\dnote}[1]{}
\newcommand{\tnote}[1]{}

\newcommand{\href}[2]{#2}
\newcommand{\eprint}[1]{{\tt #1}}
\newcommand{\eprintpub}[1]{[\eprint{#1}]}

\newcommand{\bbC}{\mathbb{C}}
\newcommand{\bbR}{\mathbb{R}}
\newcommand{\bbZ}{\mathbb{Z}}

\newcommand{\IR}{\bbR}

\newcommand{\vecx}{\vec{x}}
\newcommand{\vecp}{\vec{p}}
\newcommand{{\intkp}}{{\int_{-\infty}^{\infty} \frac{dk_0}{2\pi}
\int_{-\infty}^{\infty} \frac{d^3\vecp}{(2\pi)^3}}}
\newcommand{{\inttx}}{{\int_{-\infty}^{\infty} dt
\int_{-\infty}^{\infty} d^3 \vecx}}
\newcommand{{\intsum}}{ \sum_n \int_{-\infty}^{\infty}
\frac{d^3\vecp}{(2\pi)^3}}
\newcommand{{\inttaux}}{{\int_{0}^{\beta} d\tau
\int_{-\infty}^{\infty} d^3 \vecx}}

\newcommand{\Od}{{\cal O}}

\newcommand{\im}{\mbox{Im}}
\newcommand{\re}{\mbox{Re}}
\newcommand{\sgn}{\mbox{sgn}}

\newcommand{\pabar}{\not{\!{\partial}}}

\newcommand{\pib}{\int_{periodic} \!\!\!\!\!\!\!\!\!\!d\phi \ }

\newcommand{\prodj}{\prod_{j=1}^{2n}}

\newcommand{\prodjintc}{\prodj\int_0^\beta d\tau_j
  \int_{-\infty}^{+\infty}
dx_j}

\newcommand{\sumn}{\sum_{n=0}^{\infty}}
\newcommand{\sumno}{\sum_{n=1}^{\infty}}

\newcommand{\gsim}{\raise.3ex\hbox{$>$\kern-.75em\lower1ex\hbox{$\sim$}}}
\newcommand{\lsim}{\raise.3ex\hbox{$<$\kern-.75em\lower1ex\hbox{$\sim$}}}

\newcommand{\be}{\begin{equation}}
\newcommand{\ee}{\end{equation}}
\newcommand{\ba}{\begin{eqnarray}}
\newcommand{\ea}{\end{eqnarray}}

\newcommand{\bea}{\begin{eqnarray}}
\newcommand{\eea}{\end{eqnarray}}
\newcommand{\beq}{\begin{equation}}
\newcommand{\eeq}{\end{equation}}

\newcommand{\nnel}{\nonumber \\ {}}
\newcommand{\tref}[1]{(\ref{#1})}

\newcommand{\paa}{\partial}

\newcommand{\NP}[1]{{\it Nucl.\ Phys.\ }{\bf #1}}

\newcommand{\PL}[1]{{\em Phys.\ Lett.\ }{\bf #1}}

\newcommand{\PR}[1]{{\em Phys.\ Rev.\ }{\bf #1}}
\newcommand{\PRL}[1]{{\em Phys.\ Rev.\ Lett.\ }{\bf #1}}

\newcommand{\vol}[1]{{\bf #1}}

\begin{document}

\typeout{--- Title page start ---}

\renewcommand{\thefootnote}{\fnsymbol{footnote}}

\begin{flushright}
IMPERIAL/TP/1-02/19 \\ ORSAY-LPT-02-34\\
19th April, 2002\\
\end{flushright}
\vskip 12pt

\begin{center}

{\large\bf  Transport Coefficients and Analytic Continuation in
Dual \\ 1+1 Dimensional Models at Finite Temperature}

\vskip 1.2cm
 T.S.Evans$^{a}$\footnote{E-mail: {\tt T.Evans@ic.ac.uk}},
 A.G\'{o}mez Nicola$^b$\footnote{E-mail:
{\tt gomez@fis.ucm.es}},
 R.J.Rivers$^{c}$\footnote{E-mail: {\tt
R.Rivers@ic.ac.uk}. Permanent address (a).}
 and
 D.A.Steer$^{d}$\footnote{E-mail: {\tt
Daniele.STEER@th.u-psud.fr}}\\
 \vskip 5pt \vskip 3pt {\it a})
 \href{http://theory.ic.ac.uk}{Theoretical Physics},
 Blackett Laboratory, Imperial College,\\
 Prince Consort Road, London, SW7 2BW, U.K.\\ \vskip 3pt {\it b})
Departamento de F\'{\i}sica Te\'orica II, Universidad Complutense\\
28040, Madrid, Spain \\
\vskip 3pt {\it c})
 \href{http://theory.ic.ac.uk} {Centre of Theoretical Physics}, University of Sussex, \\
Brighton, BN1 9QJ, U.K.\\ \vskip 3pt {\it d})
 Laboratoire de Physique Th\'eorique, B\^at.\ 210, Universit\'e
 Paris XI,\\ 91405 Orsay Cedex, France
\end{center}

\vskip 1.2cm

\renewcommand{\thefootnote}{\arabic{footnote}}
\setcounter{footnote}{0} \typeout{--- Main Text Start ---}

\begin{abstract}

The conductivity of a finite temperature 1+1 dimensional fermion
gas described by the massive Thirring model is shown to be related
to the retarded propagator of the dual boson sine-Gordon model.
Duality provides a natural resummation which resolves infra-red
problems, and the boson propagator can be related to the fermion
gas at non-zero temperature and chemical potential or density. In
addition, at high temperatures, we can apply  a dimensional
reduction technique to find resummed closed expressions for the
boson self-energy and relate them to the fermion conductivity.
Particular attention is paid to the discussion of analytic
continuation. The resummation implicit in duality provides a
powerful alternative to the standard diagrammatic evaluation of
transport coefficients at finite temperature.

\end{abstract}

\vspace{1cm}

\section{Introduction and motivation}

The evaluation of transport coefficients at high temperatures in
terms of Feynman diagrams in a weakly coupled theory  is  a subtle
and highly involved task \cite{jeya96,je95}. Firstly, they are
often proportional to the mean free path of the scattering
processes, which increases as coupling strength decreases.
Further, in the relevant limits of vanishing external momenta and
energy, higher loop diagrams can be as important as lower loop
diagrams if they are sufficiently infrared sensitive. The failure
of perturbation theory that both of these observations imply
requires careful and clever resummation of diagrams.

In lower dimensions we know that this resummation is a reflection
of the fact that the relevant degrees of freedom are not those of
the quasi-particles, but of dual degrees of freedom. This is well
understood in the case of Luttinger liquids, electrons close to
the  Fermi surface in one dimension (quantum wires). Such a system
\cite{lut,hal,tom} has a dual representation of its charge modes
in terms of {\it free} bosonic fields, which provide the relevant
degrees of freedom. This leads to a huge simplification in the
calculation of conductivity \cite{apel}, not easily visible (if at
all) from electron Feynman diagrams. The predictions that this
simple duality permits have been confirmed in experimental
measurements \cite{tar} of the conductance of $GaAs$ quantum
wires, although the finiteness of the system enforces a
modification \cite{stone}  of the naive picture.

In this work we explore the advantages of using this fermion-boson
duality when calculating conductivity for relativistic quantum
fields in 1+1 dimensions. The hope is that the infinite-order
non-perturbative resummation of the quasi-particle modes can be
replaced by a few terms in the series for the dual degrees of
freedom i.e. that duality does the resummation for us. Further,
the usual calculation of transport coefficients relies on linear
response theory, and the simplifications implicit in the dual
resummation suggest that we can go beyond linear response,
otherwise impossible.

As our example, we will concentrate on perhaps the simplest
non-trivial theory displaying conductivity: the Massive Thirring
(MT) fermion theory, with Minkowski Lagrangian density
\begin{equation}
{\cal{L}}_{MT} [\bar\psi,\psi] =  \bar{\psi} (i\pabar - m_0)\psi -
\frac{1}{2}g^2  j_\mu (x) j^\mu (x)  , \label{MT}
\end{equation}
where $j_\mu=(\rho,j)=\bar\psi\gamma_\mu\psi$ is the fermion
number current.

This is dual to the sine-Gordon (SG) boson model, which is
described by
\begin{equation}
 {\cal L}_{SG}[\phi] = \frac{1}{2} \paa_\mu \phi  \paa^\mu \phi -
 \frac{\alpha_0}{\lambda^2} \cos (\lambda \phi) .
\label{sglag}
\end{equation}
provided the renormalised coupling constants are identified as \ba
\frac{\lambda^2}{4\pi}&=&\frac{1}{1+g^2/\pi}, \label{equivconst1}
\\
\frac{\alpha}{\lambda^2}&=&\rho m , \label{equivconst2}
\end{eqnarray}
where $\rho$ is the renormalisation scale and $m$ the renormalised
fermion mass.

We note that, with multiplicative renormalisation, if $m_0 = 0$,
then the {\it massless} Thirring model is dual to the {\it free}
bosonic field. If we introduce a chemical potential for the
fermion field then, in the limit of no antiparticles, we recover
the Luttinger limit. However, in QFT the massless limit is
unnatural for quarks or electrons and we are obliged to be more
sophisticated.

 While 1+1 dimensions may seem unrealistic in relativistic
 quantum field theory,
 the links of our model with conformal field theory means that
these results are relevant to more exotic situations. For instance
the study of decay of quantum normal modes of a classical field in
the presence of a black hole, all in AdS space, can be linked
through AdS/CFT correspondence to linear response in a conformal
field theory in 1+1 dimensions \cite{Bir}.

Further, as a precursor to understanding the deconfinement
transition in QCD in 3+1 dimensions, the two-dimensional fermionic
Thirring model can be put in correspondence with a compact $U(1)$
gauge theory in four dimensions \cite{yoshida1}, by virtue of
Parisi-Sourlas dimensional reduction \cite{parisi}. From this
viewpoint the chiral bilinears of the MT theory correspond to the
monopole-antimonopole pairs of the four-dimensional theory, and
the monopole condensation that signals confinement has its
counterpart in the chirally broken phase of the MT theory
\cite{yoshida2}, that we discussed in detail in an earlier
exploration of this duality \cite{grs00}.

In \cite{Belg98,gs99} it has been shown that the SG/MT models are
also equivalent at finite temperature $T$ and fermion chemical
potential $\mu$. Here, we will consider the fermion response to an
electric external field (conductivity) as a working example of the
use of duality and dimensional reduction in the calculation of
transport coefficients. This amounts to an evaluation of the full
retarded {\it boson} propagator. In this way, the self-energy of
the bosons is directly related to the conductivity and charge
screening of the fermions. Furthermore, we will show how to relate
the dynamical information contained in the boson self-energy,
calculated in the imaginary-time (IT) formalism, to the static
properties of the fermion gas. In addition, we will use
dimensional reduction techniques to analyse the boson propagator.
Dimensional reduction is meant to be valid at high temperatures
compared to the fermion mass, strong fermion couplings and large
distances compared to the inverse temperature. This regime has
been shown to be very useful in \cite{grs00}, where it was used to
resum exactly the pressure and the fermion condensate at finite
$T$ and $\mu$.

The structure of the paper is the following. In section
\ref{sec:indcurr} we derive the relationship between the fermion
induced currents in linear response theory and the SG retarded
propagator. The retarded propagator can be read off from the IT
one by analytic continuation. We devote section \ref{sec:sgpropit}
to analyse the imaginary-time SG propagator. First, we write it as
an infinite sum of mass insertions and then we explore its
relevant properties and the relationship with the fermion gas at
finite chemical potential. In section \ref{sec:dr} we discuss the
high temperature limit and the dimensional reduction regime for
the SG propagator. Analytic continuation from imaginary energy to
real energy is a fraught exercise, and this model is simple enough
to show how subtle one needs to be. This will become clear in
sections \ref{sec:dr} and \ref{phys} where we will relate the
analytic continuation of the high $T$ propagator with the physical
conductivity. We will show that one needs to impose physical
conditions on the propagator in order to have a physically
meaningful answer for the transport coefficients. Many details of
the calculation have been collected in the Appendices. Appendix
\ref{appa} contains some useful results about thermal propagators
and analytic continuation, while Appendix \ref{sgpropalg} contains
some details of  the calculations performed for the SG propagator.


\section{Induced fermion currents and the boson retarded propagator}\label{sec:indcurr}

In practice, the only transport coefficient that we will study is
electrical conductivity, $\sigma$, the response of electric
currents to an electric field, since it is for this that duality
can be easily exploited. In our case, the electric field will be
an external classical field appearing as a $c$-number source in
our action. It is convenient to absorb the gauge coupling $e$ into
the gauge fields so that the covariant derivatives are $D^\mu =
\partial^\mu - i A^\mu$ and the gauge field has dimension one. The
interaction term added to the MT Lagrangian density is then of the
form
\begin{equation}
 \delta V(t)= j_\mu (x,t) A^{\mu}_{cl}(x,t),
 \label{extsource}
\end{equation}
where the external classical electro-magnetic potential is
$A^{\mu}_{cl}$ and $j_\mu=(\rho,j)=\bar\psi\gamma_\mu\psi$ is the
fermion number current. Note that using these conventions, the
gauge field kinetic term is of the form
$-F^{\mu\nu}F_{\mu_\nu}/(4e^2)$ so the electric charge, $e$, has
dimensions of energy. As a gauge choice, we will take $A_1^{cl}=0$
in what follows, so that the external electric field is
\begin{equation}
\label{EA0rel}
 E^{cl}(x,t)=-(\partial/\partial x) A_0^{cl} (x,t).
\end{equation}
with dimension two. The electric field satisfies Maxwell's
equations in 1+1 dimensions, which read
\begin{eqnarray}
e^2 j_{cl} (x,t)&=&-\dot E_{cl} (x,t)\nonumber\\
e^2 \rho_{cl} (x,t)&=&\frac{\partial E_{cl}(x,t)}{\partial x}.
\label{maxwell}
\end{eqnarray}

With these conventions the conductivity $\sigma$ is defined by
\begin{equation}
 \label{cond}
  j = \sigma E,
\end{equation}
where $j$ is the current density and $E$ is the electric field.
The plasma is isotropic and
homogeneous so the conductivity is a simple scalar.

In defining conductivity, we are implying that the response of the
plasma to an electric field is simply linear in that field.  This
weak field limit is calculated in quantum field theory using the
theory of linear response (see \cite{kapusta,lebellac}). In terms
of the additional interaction term, $\delta V$, of
\tref{extsource}, the terms linear in electric field come from the
first term of a perturbative expansion in powers of  $\delta V$.
From linear response theory
 this is then:
\begin{equation}\label{lr1}
 \langle \langle \delta^{(1)} j_\mu (x,t) \rangle \rangle  =
  -i\int_{t_0}^{\infty} dt' \int_{-\infty}^{\infty} dx'
  A_0^{cl}(x',t')
  \langle \langle [j_\mu (x,t),j_0 (x',t')]\rangle \rangle \theta
  (t-t').
\end{equation}
The $\delta j_\mu (x,t)$ is the difference between the currents
with and without the external perturbation.  The $\delta^{(1)}$
indicates that it is the term linear in $\delta V$, and hence
linear in the electric field. $\langle \langle \cdot \rangle
\rangle$ stands for a thermal average taken with respect to an
equilibrium density matrix based on the unperturbed Hamiltonian
given at time earlier than $t_0 \in \mathbb{R}$. The perturbation,
$\delta V$ is assumed to be switched on only after $t_0$ and we
are looking at the induced current perturbation at time
$t\in\mathbb{R}>t_0$.

For the MT model of (\ref{MT}), any thermal average involving the
 fermion current $j_\mu$ can be replaced by a boson thermal average
in  the sine-Gordon model of (\ref{sglag}). Duality means the
fermion current has an exact representation in terms of the scalar
SG field alone, namely
\begin{equation} \label{dualrel}
 j_\mu (x,t)  = \frac{\lambda}{2\pi} \epsilon_{\mu\nu}
 \partial^\nu\phi(x,t),
\end{equation}
with $\lambda$ defined through (\ref{equivconst1}). The proof of
the above statement can be found in \cite{gs99}. If the fermion
fields interact electromagnetically, then we would identify $g =
O(e)$ in the effective theory described by (\ref{MT}). We feel no
need to make this identification, and treat $g$ as an independent
parameter to be varied irrespective of $e$.

Putting this together for the current then gives
\begin{equation}\label{lr2}
\langle \langle \delta^{(1)} j_1 (x,t) \rangle \rangle  =
 +i\left(\frac{\lambda}{2\pi}\right)^2
 \int_{t_0}^{\infty} dt'
 \int_{-\infty}^{\infty} dx'  A_0^{cl}(x',t') \theta (t-t')
 \frac{\partial}{\partial t}\frac{\partial}{\partial x'}
 \langle \langle
 [\phi (x,t),\phi (x',t')]
 \rangle \rangle .
\end{equation}
Using integration by parts on the $x'$ variable and using
\tref{EA0rel} gives an expression in terms of the electric field.
A resulting boundary term at spatial infinity can be ignored in
practical problems.  The time derivative can be made to act on
both expectation value and theta function by using the equal-time
commutation relation $[{\phi} (x,t),\phi (x',t)]=0$.  This then
leaves us with the linear response of the fermionic current in
terms of the applied electric field given in terms of the full
real-time retarded propagator $\Delta_R (x,t)$ of the sine-Gordon
scalar field

\begin{equation}\label{lr2b}
\langle \langle \delta^{(1)} j_1 (x,t) \rangle \rangle  =
 i\left(\frac{\lambda}{2\pi}\right)^2
 \int_{t_0}^{\infty} dt'
 \int_{-\infty}^{\infty} dx'  E^{cl}(x',t')
 \frac{\partial}{\partial t} \Delta_R (x-x',t-t'),
\end{equation}
where
\begin{equation}
\Delta_R (x-x',t-t')=\theta(t-t')\langle \langle
[\phi(x,t),\phi(x',t')]\rangle \rangle= \int_{-\infty}^{+\infty}
\frac{d\omega}{2\pi} \int_{-\infty}^{+\infty} \frac{d p}{2\pi}
e^{-i\omega t} e^{i px}
 \Delta_R (\omega,p).
\label{apdef}
\end{equation}

Taking $t_0$ to minus infinity (ignoring for now possible
subtleties in this limit \cite{TSEzm,TSEze}) allows the Fourier
transform to be taken, giving
\begin{eqnarray}
\langle \langle \delta^{(1)} j_1 (w,p) \rangle \rangle &=&
\left(\frac{\lambda}{2\pi}\right)^2 E^{cl} (\omega,p) \omega
\Delta_R (\omega,p).
\end{eqnarray}
The above result is nothing but Kubo's formula \cite{lebellac} for
the SG/MT system. Thus, the {\it conductivity} as defined in
\tref{cond} is given in terms of the SG retarded propagator as
(note that $\langle\langle j \rangle\rangle$=0 without external
fields)
\begin{equation}
 \sigma (\omega,p)=\left(\frac{\lambda}{2\pi}\right)^2
\omega \Delta_R (\omega,p). \label{cond1}
\end{equation}
Proceeding exactly in the same way for the zero component of the
current yields the {\it induced charge density}:
\begin{equation}
\langle \langle \delta^{(1)} j_0 (w,p) \rangle
\rangle=\left(\frac{\lambda}{2\pi}\right)^2 E^{cl} (\omega,p) p
\Delta_R (\omega,p). \label{cond0}
\end{equation}
So we see that all we need for the linear response to an electric
field is the full retarded sine-Gordon scalar propagator at real
Minkowski energies, which we shall analyse in detail in next
sections.

Note that:

\begin{itemize}

\item The induced current is conserved, $\partial_\mu \delta^{(1)} j^\mu=0$
  as it should.

\item In the {\it free} case, the retarded propagator is
  $T$-independent (see Appendix \ref{appa}) and so are the fermion
  conductivity and induced charge density.

\end{itemize}

We conclude this section by considering briefly the case when the
bosonic degrees of freedom are those of a free field.

\subsection{Free bosonic modes}
\label{fbm}

We see from the duality conditions that the {\it massless} ($m_0 =
0$) Thirring model is dual to the {\it massless and free} bosonic
field, for all couplings $g^2$. In this case the conductivity is
given by
\begin{equation}
 \sigma (\omega,p)=\left(\frac{\lambda}{2\pi}\right)^2
\omega \Delta^0_R (\omega,p) \label{cond10},
\end{equation}
where $\Delta^0_R (\omega,p)$ is the retarded propagator for the
massless free field.

In this, and subsequent sections, it is convenient to keep a small
nonzero mass $\mu_0$ for the scalar field and in the end we will
take $\mu_0\rightarrow 0^+$. We shall discuss retarded propagators
in great detail later. For the moment it is sufficient to quote
the result
\begin{equation}
\Delta_R^0(\omega,p) = \frac{i}{(\omega+i\epsilon)^2 - p^2 -
\mu_0^2}.
\end{equation}
We recover the conductivity from (\ref{cond10}) as
\begin{equation}
 \sigma (\omega,p)=\left(\frac{\lambda}{2\pi}\right)^2
\frac{i\omega}{(\omega+i\epsilon)^2 - p^2 - \mu_0^2}
 \label{cond100},
\end{equation}
In particular, the real part of the conductivity for constant
applied field is (after taking $\mu_0\rightarrow 0^+$)
\begin{equation}
 \sigma (\omega,0)=\pi \left(\frac{\lambda}{2\pi}\right)^2
\delta(\omega) = \frac{1}{1+g^2/\pi}\,\delta(\omega).
 \label{realcond}
\end{equation}
We shall see in section \ref{phys} that, if we apply a static
field only to a finite part of the system, then the {\it
conductance} is
\begin{equation}
G = \frac{1}{2}\left(\frac{\lambda}{2\pi}\right)^2 =
\frac{1}{2\pi(1+g^2/\pi)}. \label{G}
\end{equation}
Suppose we had not appreciated that the massless Thirring model
was dual to a free bosonic theory. We would then have attempted to
calculate the conductivity (or conductance) as a series expansion
in $g$, using the bilinear fermionic forms for the $j_{\mu}$
directly in (\ref{lr1}). The first term in the interaction picture
expansion is the simple one-loop term, which gives \cite{fenton}
\begin{equation}
 \sigma (\omega,0)=\delta(\omega) \quad \mbox{or} \quad  G = \frac{1}{2\pi}.
 \label{realcond0}
\end{equation}

We now see the power of duality in resumming the series in the
fermion coupling constant $g$. Further, the reduction in the
conductance due to the presence of repulsive interactions
($g^2>0$) has an exact counterpart in Luttinger liquids
\cite{stone}.

In subsequent sections we see how duality aids resummation for the
massive fermion theory, through the dual sine-Gordon theory.


\section{The sine-Gordon  propagator in imaginary time}\label{sec:sgpropit}

We will adopt a similar approach to that of \cite{gs99,grs00} in
expanding in fermion mass about our results for the {\it massless}
fermionic theory and its dual bosonic free theory counterpart.

As before, it is necessary that the bosonic calculations are
moderated by a small nonzero mass $\mu_0$.
In \cite{gs99,grs00} it is shown that observables such as the
pressure and the quark condensate are $\mu_0$-independent and
hence infrared finite. This is true also for correlators evaluated
at different space-time points \cite{gs99,grs00} as long as they
involve {\it at least one derivative} of the scalar field. The
conductivity as it stands in (\ref{cond1}) should be also $\mu_0$
independent, because it is proportional to the time derivative of
the two-point function, and so is $p\Delta_R (\omega,p)$ in
(\ref{cond0}).

The {\it imaginary time} propagator is obtained by differentiating
twice the generating functional
\begin{eqnarray}
\Delta_T(x,\tau)&=&\frac{1}{Z_{SG} (T)} \left.\frac{\delta}{\delta
  J(x,\tau)}\frac{\delta}{\delta J(0)}
  Z_{SG}[J;T]\right\vert_{J=0},
 \nonumber\\
 Z_{SG}[J;T]&=& N_\beta
  \pib \exp\left\{-\int_0^\beta d\tau \int_{-\infty}^{\infty}\left[
  {\cal L}_{SG}[\phi]+J(x)\phi(x)\right]\right\},
 \label{sggf}
\end{eqnarray}
where $\tau\in[0,\beta]$, $\beta=T^{-1}$ and $Z_{SG}
(T)=Z_{SG}[0;T]$ is the SG partition function, which coincides
with the MT model one $Z_{SG} (T)=Z_{MT} (T)$ as showed in
 \cite{Belg98,gs99}.

The SG generating functional (\ref{sggf}) can be given explicitly
as a power series in $\alpha$  (see equation (3.16) in
 \cite{gs99}). With $\alpha/\lambda^2 = m\rho$, this is an
expansion in fermion mass about the free boson theory. Setting
$J=0$ one has the SG partition function:

\begin{eqnarray}
 \frac{Z_{SG} (J=0, T)}{Z_0^B (T)}
 =1+
 \sumno
 \left(\frac{1}{n!}\right)^2
 \left[ \frac{\alpha}{2\lambda^2}
 \left(\frac{T} {\rho}\right)^{\lambda^2/4\pi} \right]^{2n}
 \nonumber\\
 \times
 \left( \prodjintc \right)
 \left( \prod_{j=2}^{2n}\prod_{k=1}^{j-1} \left[
 Q^2(x_j-x_k,\tau_j-\tau_k)
 \right]^{\epsilon_j\epsilon_k \lambda^2/4\pi} \right),
 \label{ZSGexplicit}
\end{eqnarray}
where $Z_0^B (T)$ is the free boson partition function (equal to
$Z_0^F (T)$, the free fermion partition function in 1+1
dimensions), and
\begin{equation}
\epsilon_j=\left\{ \begin{array}{cl} +&j=1,\dots,n\\
                                     -&j=n+1,\dots,2n
\end{array}\right. .
\label{epsdef}
\end{equation} The $Q^2$ function is given by
\begin{equation}Q^2(x,\tau)
 = \sinh(\frac{\pi (x +i\tau )}{\beta})
 \sinh(\frac{\pi (x- i\tau )}{\beta}),
 \label{qs}
\end{equation}
and $\Delta_T^0(x,\tau)$ is the free boson IT propagator in the
$\mu_0\rightarrow 0^+$ limit \cite{Belg98,gs99}:
\begin{equation}
 \Delta_T^0(x,\tau) = -\frac{1}{4\pi} \ln \mu_0^2\beta^2
 Q^2(x,\tau).
\label{DeltaT0}
\end{equation}

It can be useful to visualise  expressions like
(\ref{ZSGexplicit}) in terms of diagrams representing the terms in
$\alpha$ expansions such as \tref{ZSGexplicit} or
\tref{sgitprop1}. Each expression has $2n$ integrations over $2n$
space-time coordinates, each represented by a vertex.  They  split
into two types according to the choice \tref{epsdef}.  The $n$
vertices associated with $\epsilon_j=+1$ and integrals over the
$(x_1,\tau_1)$ to $(x_n,\tau_n)$ coordinates we will denote by a
closed circle. The remaining $n$ vertices have $\epsilon_j=-1$ and
integrals over the $(x_{n+1},\tau_{n+1})$ to $(x_{2n},\tau_{2n})$
coordinates, and we use an open circle.  It is a crucial rule that
there are always as many open as closed circle vertices as a
result of a superselection rule, the derivation of which was a key
element of \cite{gs99}. This rule is nothing but taking the
$\mu_0\rightarrow 0^+$ limit so that physical quantities are IR
finite. There is also a factor of $\alpha / \lambda^2 = m\rho$ per
vertex. Note that in terms of the fundamental $\phi$ field, tag
corrections ($\Delta(0)$ factors) of the $\phi$ field vertices
(when thinking in terms of an expansion of the $\cos ( \lambda
\phi)$) have been absorbed leading to a $(T/
\rho)^{\lambda^2/4\pi}$ renormalisation factor for each $\alpha$,
e.g.\ through the use of thermal normal ordering \cite{gs99}.
Thus, each of these vertices represents a factor of
\begin{equation}
 \int_0^\beta d\tau_j \int_{-\infty}^{\infty} dx_j
 \frac{\alpha(T)}{2\lambda^2}, \;\;\;
 \alpha(T) := \alpha \left(\frac{T} {\rho}\right)^{\lambda^2/4\pi}
 = \alpha_0 \left(\frac{T^2} {\Lambda^2}\right)^{\lambda^2/8\pi},
 \label{alphafactor}
\end{equation}
with the $\epsilon$ factors associated with each vertex modifying
the ``propagators'' attached to them. The second form for
$\alpha(T)$ is in terms of the bare coupling and the UV cutoff
$\Lambda$, as given in \cite{gs99}.

In every diagram each of the $2n$ vertices are connected once to
every other vertex by a ``propagator'', drawn as a double line.
The double line linking the $j$-th and $k$-th vertices represents
a factor of $(Q^{\lambda^2/2\pi})^{\epsilon_i\epsilon_j}$ so that
lines connecting one open and one closed vertices contribute to
the denominator while those connecting vertices of the same type
are part of the numerator.  Without the factors of $\epsilon$ from
the vertices, we have

\begin{eqnarray}
\left[Q^2 (x,\tau)\right]^{-\lambda^2/4\pi} &=&
\begin{picture}(40,10)(0,-4)
 \put(0,1){\line(1,0){40}}
 \put(0,-1){\line(1,0){40}}
\end{picture}
\sim
 \exp[\lambda^2 \Delta_0]
\nonumber\\
\hfill\nonumber\\
 &\sim&
\begin{picture}(340,10)
\put(1,1){\circle*{3}} \put(5,-3){$+$}
\put(20,-1){$\displaystyle\lambda^2$} \put(35,0){\line(1,0){40}}
\put(90,-3){$+$}\put(110,0) {$\displaystyle\frac{1}{2}\lambda^4$}
\put(150,0){\circle{25}} \put(180,0){$+$}
\put(200,0){$\displaystyle\frac{1}{3!}\lambda^6$}
\put(240,0){\circle{25}} \put(228,0){\line(1,0){24}}
\put(270,0){$+ \ \ \dots$}
\end{picture}
 \label{dldef}
\end{eqnarray}

\vspace*{.3cm}

In terms of the $\phi$ field each $Q$ propagator line represents
many different $\phi$ diagrams with $\phi$ vertices of all even
orders coming from the expansion of the $\cos$ factors in the
Lagrangian. However a $\lambda$ expansion is not a good way to
appreciate the result as a whole because of IR problems. In fact,
note that the original expression for the partition function (and
the same will happen for the propagator) was $\mu_0$ independent,
while every single diagram in (\ref{dldef}) depends on $\mu_0$,
giving rise to divergent terms as $\mu_0\rightarrow 0^+$, which is
the limit where the identification $Q^2\sim \exp [-4\pi \Delta_0]$
done in (\ref{dldef}) is valid. We will therefore keep the double
line resumed propagator and we will discuss in section
\ref{sec:dr} its interpretation in the high temperature limit.

Since the $Q$ functions \tref{qs} are invariant under $x_j
\Leftrightarrow x_k$ then no sense of direction need be assigned
to these double line propagators, and there is a symmetry between
diagrams related by an exchange of open and closed vertices.

Summing up, the vacuum diagrams can be represented by sums over
diagrams of the type illustrated
\begin{eqnarray}
 \frac{Z_{SG} (J=0, T)}{Z_0^B (T)}
 &=&
 \sumno
\raisebox{-1cm}{\includegraphics{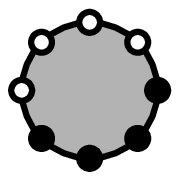}}
 \\
 &=&
 \raisebox{-4mm}{\includegraphics{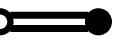}}
 +
 \includegraphics{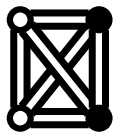} + \ldots +
{\includegraphics{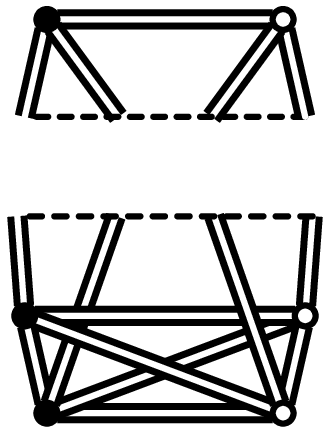}} + \ldots
\end{eqnarray}

Note that the interaction was based on a $\cos ( \lambda \phi)$
and not a $\cos ( \lambda \phi) - 1$ factor, so there is a
physically irrelevant shift in the action and resulting partition
function. Also note that though this is a calculation of the
partition function $Z$ and not $\ln Z$, the diagrams are
connected. Finally, we remark that the above diagrams are in
position, not momentum, space and there are only $2n-1$
independent integration variables in each term in the sum due to
translation invariance, as it is emphasized in Appendix
\ref{sgpropalg} for the case of the propagator.

Following similar steps as in \cite{gs99} for the partition
function, one can write the full IT propagator as an infinite
series in the renormalised $\alpha$ \tref{alphafactor}. Again the
$\mu_0 \rightarrow 0^+ $  IR limit must be dealt with properly,
but in the end we can express the result in the same terms as the
partition function with a couple of extra rules. First, as in
normal perturbation theory, one will `pull out' two free $\phi$
propagators $\Delta^0_T \sim \ln (-Q^2/4\pi) $ (see Appendix
\ref{sgpropalg} for details) of \tref{DeltaT0}, which we will
denote with a single line. These connect to any of the open and
closed vertices appearing in the partition function, but for each
additional single $\phi$ field line attached to a vertex $j$ one
multiplies by a factor of $-i \epsilon_j \lambda/(4 \pi)$.
However, a key result in this model is that one has only one or
two free $\phi$ propagators contributing to the full propagator.
One does {\em not} get chains of self-energy insertions separated
by a single free $\phi$ propagator. We will comment about this
observation in section \ref{sec:dr}.

Thus, the full propagator has only four types of contribution, a
single free $\phi$ propagator and three types of diagrams with
interactions.
\begin{eqnarray}
 \Delta_T(x,\tau)
 &=&
 \Delta_T^0(x,\tau) +
 \frac{Z_0^B (T)}{Z_{SG} (T)}
 \sumno
 \Gamma^{(2n)}_2 (x,\tau ),
 \\
 \Gamma_{2}^{(2n)} (x,\tau)
 &=&
   \Gamma_{2\pm}^{(2n)} (x,\tau)
 + \Gamma_{2\pm\mp}^{(2n)} (x,\tau)
 + \Gamma_{2\pm\pm}^{(2n)} (x,\tau).
 \label{sgitprop1b}
\end{eqnarray}
The three types of interacting terms are given in terms of a
similar formula, differing only in how the two free propagators
are convolved with the $Q^2$ propagators.  Writing those terms as
$ \left\{ \Delta \Delta \right\}_A$ ($A=\pm,\pm\mp,\pm\pm$) the
general expression is of the form (see details of the derivation
in Appendix \ref{sgpropalg}):
\begin{eqnarray}
\Gamma_{2A}^{(2n)} (x,\tau )
 &=&
 -\left(\frac{\lambda}{4\pi}\right)^2
 \left(\frac{1}{n!}\right)^2
 \left[ \frac{\alpha}{2\lambda^2}
 \left(\frac{T} {\rho}\right)^{\lambda^2/4\pi} \right]^{2n}
 \left( \prodjintc \right)
 \nonumber\\
 &&  \times
 \left\{ \Delta \Delta \right\}_A
 \left( \prod_{j=2}^{2n}\prod_{k=1}^{j-1} \left[
 Q^2(x_j-x_k,\tau_j-\tau_k)
 \right]^{\epsilon_j\epsilon_k \lambda^2/4\pi} \right).
 \label{sgitprop1}
\end{eqnarray}
where
\begin{eqnarray}
 \left\{ \Delta \Delta \right\}_{\pm}
 &=&
   2n \Delta^0_T(x_1,\tau_1)
 \Delta^0_T(x_1-x,\tau_1-\tau),
\\
 \left\{ \Delta \Delta \right\}_{\pm\mp}
 &=&
   -2n^2 \Delta^0_T(x_1,\tau_1)
 \Delta^0_T(x_{2n} - x , \tau_{2n} - \tau),
\\
 \left\{ \Delta \Delta \right\}_{ \pm \pm }
 &=&
  2n(n-1)\Delta^0_T(x_1,\tau_1)
 \Delta^0_T(x_2-x,\tau_2-\tau) .
\end{eqnarray}
Note that we have exploited the space-time translation invariance
of an equilibrium system as well as the symmetry between open and
closed vertices to choose specific space-time variables for the
free propagators with the remaining equivalent choices being
accounted for by simple combinatorial prefactors.  Such factors
are obvious when we write these in our diagrammatic notation as
\begin{eqnarray}
   \Gamma_{2\pm}^{(2n)}   (x,\tau)
  &=&
 \begin{picture}(100,100)
{\includegraphics{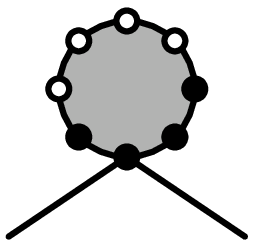}}
\end{picture}
+ \left( \begin{picture}(45,5) \put(2,3){\circle*{3}}
 \put(10,0){$\longleftrightarrow$}  \put(40,3){\circle{3}}
 \end{picture}\right)
\\
&\approx &
 {\includegraphics{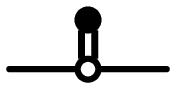}}
+ {\includegraphics{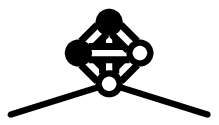}} + \left( \begin{picture}(45,5)
\put(2,3){\circle*{3}}
 \put(10,0){$\longleftrightarrow$}  \put(40,3){\circle{3}}
 \end{picture}\right)
 + \ldots
 \label{pm}
\\
 \Gamma_{2\pm\mp}^{(2n)}  (x,\tau)
 &=&
{\includegraphics{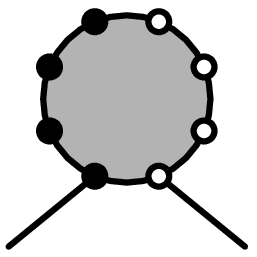}} +\left( \begin{picture}(45,5)
\put(2,3){\circle*{3}}
 \put(10,0){$\longleftrightarrow$}  \put(40,3){\circle{3}}
 \end{picture}\right)
\\
& \approx &
{\includegraphics{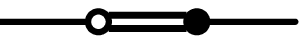}}
 +
{\includegraphics{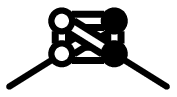}} + \left( \begin{picture}(45,5)
\put(2,3){\circle*{3}}
 \put(10,0){$\longleftrightarrow$}  \put(40,3){\circle{3}}
 \end{picture}\right)
+ \ldots \label{pmmp}
 \\
 \Gamma_{2\pm\pm}^{(2n)}  (x,\tau)
 &=&
 {\includegraphics{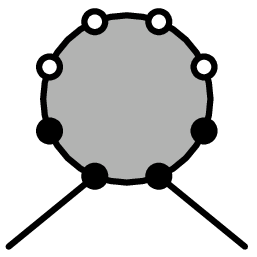}}
+ \left( \begin{picture}(45,5) \put(2,3){\circle*{3}}
 \put(10,0){$\longleftrightarrow$}  \put(40,3){\circle{3}}
 \end{picture}\right)
 \approx
 {\includegraphics{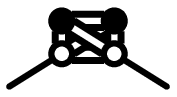}}
+ \left( \begin{picture}(45,5) \put(2,3){\circle*{3}}
 \put(10,0){$\longleftrightarrow$}  \put(40,3){\circle{3}}
 \end{picture}\right)
 + \ldots
\label{pmpm}
\end{eqnarray}

The last two diagrams above can be combined in interesting ways,
exploiting the fact that they differ only by the exchange of an
open for a closed vertex on an external leg, with a compensating
change on an internal vertex.

It is not difficult to check that all the integrals appearing in
the above expressions for the propagator are finite in the
infra-red, i.e, for large spatial distances, following the same
arguments as in \cite{gs99} for the partition function. In
addition (see Appendix \ref{sgpropalg} for details) the only
$\mu_0$-dependence in (\ref{sgitprop1}) appears in the free part
$\Delta^0 (x,\tau)$, in the $\mu_0\rightarrow 0^+$ limit. In
Appendix \ref{sgpropalg} it is also shown that the integrals in
(\ref{sgitprop1}) are UV (short distances) finite provided
$\lambda^2<4\pi$. We shall restrict here to $\lambda^2<4\pi$,
where the theory is superrenormalisable (see comments in
 \cite{gs99,grs00} in this respect). On the other hand, the free
propagator is UV divergent. Therefore, all the IR and UV
divergences are in the free part of the full IT  propagator for
$\lambda^2<4\pi$. Hence, every term in the $\alpha$-expansion in
(\ref{sgitprop1}) is infra-red (and UV) finite, which indicates
that this is the appropriate expansion we should look at to avoid
the problems mentioned in the introduction, and related here to
the {\it naive} $\lambda$ expansion.

We also remark that (\ref{sgitprop1}) is independent of the scale
$\rho$, since the explicit dependence is exactly compensated by
the implicit scale dependence in $\alpha$ (see
 \cite{gs99} for more details).
 A conventional choice of scale is $\rho=m$, the mass of the
fermion. This is particularly interesting at high temperatures
$T\gg m$ where the $\alpha$ expansion in (\ref{sgitprop1}),
originally seen as an expansion in fermion mass $m^2$, becomes
effectively a high temperature series in
$(m^2/T^2)^{1-\lambda^2/8\pi}$ \cite{grs00} (see section
\ref{sec:dr}). Thus, at high temperatures, we can consistently
truncate the series (\ref{pm}), (\ref{pmmp}) and (\ref{pmpm}),
although we shall not do so yet. Notice that we {\it fix} the mass
of the fermion, meaning in particular that $\alpha\rightarrow 0^+$
in the $\lambda\rightarrow 0^+$ limit. In other words, even in
that limit we do not recover a massive free boson theory, as one
would naively expect from the lagrangian (\ref{sglag}), by
expanding to lowest order in $\lambda$.

 Now let us  Fourier transform the propagator (\ref{sgitprop1})
to momentum space $\Delta(i\omega_n,p)$  with {\it discrete}
frequencies $\omega_n$. We can write the result as:
\begin{equation}
\Delta_T(i\omega_n,p)= \Delta_T^0(i\omega_n,p) \left[1+\Sigma_T
(i\omega_n,p)\Delta_T^0(i\omega_n,p)\right], \label{sgap1}
\end{equation}
where $\Delta_T^0(i\omega_n,p)$ is the free propagator in momentum
space:
\begin{equation}
\Delta_T^0(i\omega_n,p)=\frac{1}{\omega_n^2+p^2+\mu_0^2} .
\end{equation}
where we have kept the dependence with the small $\mu_0$ for
reasons to become clear below.

Note that $\Sigma$ is {\em not} a self-energy in the usual sense,
but rather it is just the truncated full propagator.  It is 1PI
but the full propagator is {\em not} an infinite sum of such
insertions. The conventional Schwinger-Dyson form for the
propagator in momentum space reads schematically
$\Delta_T=\Delta_T^0 (1-\Pi_T \Delta)$ where $\Pi_T$ is the true
self-energy. Therefore, the imaginary-time self-energy is related
to  $\Sigma_T$ as
\begin{equation}\Pi_T (i\omega_n,p) = -\frac{\Sigma_T
  (i\omega_n,p)}{1+\Sigma_T(i\omega_n,p)\Delta_T^0(i\omega_n,p)}.
\label{sedef}
\end{equation}

In more detail, the non-trivial part, the $\Sigma_T
(i\omega_n,p)$, can be written as (see details of the calculation
in Appendix \ref{sgpropalg}) :

\begin{equation} \Sigma_T (i\omega_n,p)
 = \frac{\lambda^2}{4\pi}  \frac{Z_0^B (T)}{Z_{SG} (T)}
\sumno \left(\frac{1}{n!}\right)^2
\left[\frac{\alpha}{2\lambda^2}\left(\frac{T}
{\rho}\right)^{\lambda^2/4\pi}\right]^{2n} \Sigma_T^n
 (i\omega_n,p),
\label{se1}
\end{equation}
with

\ba
 \Sigma_T^n (i\omega_m,p)&=&-8\pi \prod_{j=2}^{2n}\int_0^\beta d\tau'_j
  \int_{-\infty}^{+\infty}dx'_j \prod_{1 \leq k < j \leq 2n}
 \left[ Q^2(\sum_{l=k+1}^j x'_l , \sum_{l=k+1}^{j}\tau'_l)
        \right]^{\epsilon_j\epsilon_k \lambda^2/4\pi}
 \nonumber
 \\
 &\times&
 \left\{n+n(n-1)e^{i\omega_m
 \tau'_2} e^{-ipx'_2}-n^2 e^{i\omega_m\sum_{l=2}^{2n} \tau'_l}
 e^{-ip\sum_{l=2}^{2n} x'_l}\right\}.
 \label{itsefac}
\end{eqnarray}

Note that, from (\ref{itsefac}) it is clear that $\Sigma_T
(0,0)=0$ and $\Sigma_T (i\omega_m,p)=\Sigma_T
(-i\omega_m,p)=\Sigma_T
  (i\omega_m,-p)$. Besides, from the asymptotic behaviour of the $Q^2$ variables as
  $x'_j\rightarrow \pm \infty$ it is not difficult to see that
 the  Taylor series of $\Sigma_T (i\omega_m,p)$ around $p=0$ for fixed
 $\omega_m$ is well defined, so that  we can write

\begin{equation}\Sigma_T (\omega_m=0,p)=\sum_{k=1}^\infty a_k (T) p^{2k}
\label{sigmm0fexp}. \end{equation}

In addition, note that, from (\ref{sgap2}) one has  $[\Sigma_T
\Delta_T^0]
  (0,0)=0$ (since $\sum_{k=1}^{2n} \epsilon_k =0$). To
  make this result compatible with (\ref{sigmm0fexp}), one has
   to keep the small mass in the {\it
  free}
  propagator $\mu_0\neq 0$ till the very end of the calculation, since
\be
[\Sigma_T \Delta_T^0]  (\omega_m=0,p)=\frac{a_1 p^2+\Od
  (p^4)}{p^2+\mu_0^2},
\ee so that $[\Sigma_T \Delta_T^0]  (\omega_m=0,0)=0$ if
$\mu_0\neq 0$, and so we will do in the following.

Here we will be mainly interested in the analytic continuation of
 $\Sigma_T (i\omega_n,p)$ to real frequencies $\omega$ and its
 behaviour for small frequencies and long wavelengths (small $p$).
 Before discussing the analytic continuation, we will derive an
 interesting  relationship between the lowest order coefficient $a_1$ in
 the momentum expansion (\ref{sigmm0fexp}) and the MT model at
 finite density.
 It is a good example of
 how the use of duality can yield interesting and unexpected
 connections between the boson self-energy and the fermion gas.

 \subsection{The lowest order boson self-energy and
 the fermion charge density}
\label{sec:low}

 From (\ref{sigmm0fexp}) and
(\ref{itsefac}) we have

\ba
 a_1^n (T)&=&4\pi \prod_{j=2}^{2n}\int_0^\beta d\tau'_j
  \int_{-\infty}^{+\infty}dx'_j \prod_{1 \leq k < j \leq 2n}\left[ Q^2(\sum_{l=k+1}^j
x'_l,\sum_{l=k+1}^j \tau'_l)\right]^{\epsilon_j\epsilon_k
\lambda^2/4\pi} \nonumber\\ &\times& \left\{n(n-1) [x'_2]^2- n^2
\left[\sum_{l=2}^{2n} x'_l\right]^2\right\}. \label{itsefaclo}
\end{eqnarray}

Here, $a_1^n (T)$ denotes the $\Od(p^2)$ coefficient of
$\Sigma_T^n (0,p)$. Let us now perform back the change of
variables (\ref{cov})-(\ref{invcov}) and put the system on a
finite length $L$, so that $\int_{-\infty}^{\infty}dx_j\rightarrow
\int_{-L/2}^{L/2} dx_j$. In the end we shall take
$L\rightarrow\infty$. Thus, (\ref{itsefaclo}) becomes

\ba  a_1^n (T)&=& \frac{4\pi}{\beta L}\prod_{j=1}^{2n}
\int_0^\beta d\tau_j
  \int_{-L/2}^{L/2}dx_j \prod_{1 \leq k < j \leq 2n}\left[
  Q^2(x_j-x_k,\tau_j-\tau_k) \right]^{\epsilon_j\epsilon_k
\lambda^2/4\pi} \nonumber\\ &\times& \left\{n(n-1)
\left(x_2-x_1\right)^2- n^2 \left(x_{2n}-x_1\right)^2\right\}.
\label{itsecomp} \end{eqnarray}

At this point, let us recall that the MT model partition function
at {\it nonzero chemical potential} $\mu$ can be written as
\cite{gs99,grs00}

\ba
Z_{MT} (T,\mu)=Z_0^F (T) \exp\left[\beta L
  \frac{\mu^2}{2(\pi+g^2)}\right] \sumn \left(\frac{1}{n!}\right)^2
\left[\frac{\alpha}{2\lambda^2}\left(\frac{T}
{\rho}\right)^{\lambda^2/4\pi}\right]^{2n} F_{2n} (T,\mu),
\label{pfmu} \end{eqnarray} with

\begin{equation} F_{2n} (T,\mu)=\prod_{j=1}^{2n}
\int_0^\beta d\tau_j
  \int_{-L/2}^{L/2}dx_j \prod_{1 \leq k < j \leq 2n}\left[
  Q^2(x_j-x_k,\tau_j-\tau_k) \right]^{\epsilon_j\epsilon_k
\lambda^2/4\pi}
\exp\left[i\mu\frac{\lambda^2}{4\pi}\sum_{j=1}^{2n}\epsilon_j
x_j\right], \label{f2n} \end{equation}
 $Z_0^F (T)=Z_0^B (T)$ being the free fermion (or boson) partition function.

Consider now

\ba \left.\frac{\partial^2}{\partial \mu^2} F_{2n}
(T,\mu)\right\vert_{\mu=0}&=&
-2\left(\frac{\lambda^2}{4\pi}\right)^2 \prod_{j=1}^{2n}
\int_0^\beta d\tau_j
  \int_{-L/2}^{L/2}dx_j \prod_{1 \leq k < j \leq 2n}\left[
  Q^2(x_j-x_k,\tau_j-\tau_k) \right]^{\epsilon_j\epsilon_k
\lambda^2/4\pi}\nonumber\\
&\times& \left\{ n x_1^2 + n(n-1) x_1 x_2 - n^2 x_1 x_{2n}
\right\},
\end{eqnarray}
where we have relabelled the $x_j$ variables as explained in
Appendix \ref{sgpropalg} for similar calculations. Further
relabelling allows us to replace in the above integral:

\ba
 n x_1^2 + n(n-1) x_1 x_2 - n^2 x_1 x_{2n}&\rightarrow&
\frac{1}{2} n(1-n) \left(x_1^2+x_2^2-2 x_1 x_2\right)
+\frac{n^2}{2} \left(x_1^2 + x_{2n}^2 -2 x_1 x_{2n}\right)
\nonumber\\
&=& -\frac{1}{2}\left[ n (n-1) (x_2-x_1)^2-n^2
(x_{2n}-x_1)^2\right],
\end{eqnarray}
and therefore, by comparing with (\ref{itsecomp}) we find

\be  a_1^n (T)= \frac{4\pi}{\beta
L}\left(\frac{4\pi}{\lambda^2}\right)^2
 \left.\frac{\partial^2}{\partial \mu^2} F_{2n}
   (T,\mu) \right\vert_{\mu=0}.
\end{equation}

Let us turn this relationship into one involving physical
observables. The pressure of the MT model gas is \beq P_{MT}
(T,\mu)=\lim_{L\rightarrow\infty} \frac{1}{\beta L} \ln Z_{MT}
(T,\mu), \end{equation} and the fermion charge density is \beq
\rho_{MT} (T,\mu)=\frac{\partial}{\partial\mu} P_{MT} (T,\mu)=
\frac{\mu}{\pi+g^2}+\rho_C (T,\mu), \end{equation} where the first
term in the r.h.s. is the {\it massless} Thirring model charge
density \cite{AlvGom98} and the second one is given in
 \cite{grs00}. Then, collecting our previous results and recalling
that   $Z_{MT} (T,\mu=0)=Z_{SG} (T)$ \cite{gs99}  we find for the
lowest order boson self-energy: \beq
 a_1 (T) =\frac{16\pi^2}{\lambda^2} \left.
 \frac{\partial}{\partial\mu} \rho_C (T,\mu)\right\vert_{\mu=0}.
 \label{exactcd}
\end{equation}
This equation provides an exact relationship between the leading
low-energy boson self-energy and the fermion charge density
subtracting the massless part, as a direct consequence of the
fermion-boson thermal duality.

\section{High temperature approximate propagator}\label{sec:dr}

\subsection{Static observables}

Before looking at dynamical quantities at high temperature, such
as conductivities, it is worthwhile recalling how one can study
time-independent quantities in the SG/MT system.  Just as in
other models in other space-time dimensions, the large temperature
means that the Euclidean time dimension becomes very small and one
expects {\em dimensional reduction} (DR) to occur
 \cite{gi80,appi81,ka96}.

For the SG/MT model, the nature of dimensional reduction and its
application to static quantities was studied in \cite{grs00}. The
key observation was that the $Q^2$ variables of \tref{qs}, the
essential building blocks of all the exact SG/MT expressions, have
a large-distance limit of
\begin{equation}
 Q^2(x,\tau)\longrightarrow \frac{1}{4}\exp{\frac{2\pi \vert x\vert}{\beta}}
  \qquad \mbox{for} \quad
  \forall \;  \frac{|\tau|}{\beta} \leq 1 \ll  \frac{|x|}{\beta} .
 \label{qasym}
\end{equation}
If the relevant distance scales for the physics of interest are
much longer than $\beta$, then we will be able to use this
approximation for the $Q^2$.  We will refer to this limit as DR,
since the dependence of $Q^2$ on imaginary time $\tau$ disappears.
For the {\em static} quantities, such as the pressure, the precise
conditions required for DR to be a good approximation were shown
in \cite{grs00} to be that either $T \gg m$ with $g^2>0$
($\lambda^2<4\pi$) or  $T \gtrsim m$ with $g^2\gg \pi$ ($\lambda
\ll 4\pi$), with  $m$ held fixed. Here, $m$ is the fermion mass at
the scale $\rho=m$.

The main utility of DR for the SG/MT system is that it allows us
to write $\alpha$-expansions such as (\ref{pfmu})-(\ref{f2n}) in
terms of a {\it classical} one-dimensional gas of $2n$ charged
particles whose positions on the line are labeled by the $x_i$,
subject to the Coulomb potential $V(x_1,x_2)\propto \vert
x_1-x_2\vert$. This system was solved exactly a long time ago
 \cite{lenard1,lenard2}, so that in the DR regime we can resum the
$\alpha$ series and obtain exact results for the thermal SG/MT
system \cite{grs00}. In particular the pressure is a static
quantity and thus the phases of  SG/MT can be obtained. As shown
in \cite{grs00}, one can identify a ``molecular'' phase, where
fermion condensates tend to pair  forming "molecules" (in analogy
with the behaviour of Coulomb charges in \cite{lenard1,lenard2})
which are responsible for the  chiral symmetry restoration as
$T\rightarrow\infty$. There is also a lower $T$ regime, or
"plasma" phase, where condensates pair less easily and the chiral
symmetry is broken.

For quantities with no space-time dependence, such as the
pressure, the applicability of DR is intuitively obvious and can
be made mathematically precise as in \cite{grs00}. On the other
hand, as it is also shown in \cite{grs00}, if we are dealing with
 space-dependent but static objects, such as correlators of
fields separated in space but not time, then the DR approximation
for $Q^2$ is a good one provided we work in the limits $T\gg m$,
$\lambda \ll 4\pi$ and we enforce an extra condition. That is that
any external spatial distance scale, $x$, satisfies $\vert x
\vert\gg\beta/\pi$. In terms of Fourier components this means that
the approximations can only describe `soft' physics, that is
physics of spatial momenta $p \ll T$.

Of course, in the DR regime the simple form for the $Q^2$
\tref{qasym} means we can obtain closed results for the different
expressions.  However here we wish to study {\em time} dependent
quantities, and it is not at all clear that a time-independent
approximation for the $Q^2$, such as the DR form, can have
anything to say about the study of dynamical quantities. In fact,
as we will see now, the time-independent approach amounts to
neglect all but the $n=0$ Matsubara mode. However, the $n\neq 0$
modes do play a crucial role in physical situations.

As first pointed out by Pisarski \cite{Pis}\anote{Fix reference!}
the study of dynamical quantities at high temperature in
four-dimensional weak coupled theories requires that one resums an
infinite set of diagrams as the high temperature can compensate
for the small coupling. This has been since been developed in
detail from several perspectives \cite{HTL}\anote{Fix reference!}.
The original diagrammatic viewpoint of Braaten and Pisarski
\cite{BP} is that one has to resum HTL (Hard Thermal Loops) ---
the leading temperature dependent contributions from diagrams of
arbitrary order in coupling.  This leading temperature dependence
comes only from the hard modes, where energy and momentum is at
least of order $T$. However we can also view this as producing an
effective action describing the `soft' physics in 3+1 dimensions,
that is on energy or momentum scales of order $E,p \lesssim gT$ or
less, where $g$ is a gauge coupling or equivalent. To get such an
action, the hard modes must be integrated out of the theory.  The
point is that such an effective theory can describe dynamics of
soft processes not just static quantities. Moreover, the HTL
effective action has non-trivial dependence on energy (at least
for relativistic fermion and gauge fields in 3+1 dimensions) and
so differs from the energy independent DR effective action. Though
soft-momentum dependence is described by the HTL effective action,
it is only for static quantities that it reduces to the DR action.
The results of \cite{HTL}, if not the general ideas, are specific
to 3+1 dimensions and to relativistic theories. One of our tasks
here is to study this question in our models. We will show in next
sections that one has a hierarchy of soft and hard scales similar
to the HTL one and also emerging from an effective resummation.


\subsection{The imaginary-time SG  propagator at high $T$}

We will work in the following limits: $T\gg m$ ($m$ is the fixed
fermion mass), $p\ll 2\pi T$ and $\lambda^2\ll 4\pi$ for the
reasons explained above.

In the $T\gg m$ limit, it is justified to keep only the
$\Od(\alpha^2)$ term in the $\alpha$-expansion for the IT
$\Sigma_T$ in   (\ref{se1}). We follow similar arguments as in
\cite{grs00}. That is, we shift $\tau'_j\rightarrow \beta \tau'_j$
and $x'_j\rightarrow \beta x'_j$ in (\ref{itsefac}) so that we can
write (\ref{se1}) for $\rho=m$ as

\begin{equation} \Sigma_T (i\omega_n,p)
 = T^2 \frac{\lambda^2}{4\pi}  \frac{Z_0^B (T)}{Z_{SG} (T)}
\sumno \left(\frac{1}{n!}\right)^2
\left[\frac{1}{2}\left(\frac{m^2}
{T^2}\right)^{1-\lambda^2/8\pi}\right]^{2n} \tilde\Sigma_T^n
 (2\pi i n,p/T),
 \label{acsgap2a}
\end{equation} where $\tilde\Sigma_T^n$ is a
dimensionless function obtained from $\Sigma_T^n$ in (3.12) by
taking $\beta=1$, so that it depends weakly on $T$ for small $p/T$
(it is independent of $T$ for $p=0$). If we do the same with the
partition function $Z_{SG} (T)$ as discussed in \cite{grs00},
(\ref{acsgap2a})  yields a well-defined expansion in $m/T$
(remember that $\lambda^2/4\pi<1$). The advantage of using this
expansion instead of the naive $\lambda$ expansion is that it is
free of the problems mentioned in the introduction, like the bad
IR behaviour.

Hence,  taking
 the leading order in the high-$T$ expansion gives:

\begin{equation}\Sigma_T (i\omega_n,p)\approx \frac{\lambda^2}{2}m^2 T^2
\left(\frac{m^2}{T^2}\right)^{1-\lambda^2/4\pi}
\left[h(i\omega_n,p)-h(0,0)\right], \label{achprev}
\end{equation}
with

\begin{equation}h(i\omega_n,p)=\int_0^\beta d\tau \int_{-\infty}^{\infty} dx \
e^{i\omega_n
  \tau} e^{-i p x} \left[Q^2 (x,\tau)\right]^{-\lambda^2/4\pi}.
\label{ach}
\end{equation}

Taking into account that the full IT propagator is $\Delta_T=
\Delta_T^0 + \Sigma_T \left[\Delta_T^0\right]^2$,
 keeping only the term (\ref{achprev}) can be
interpreted diagrammatically as:
\begin{eqnarray}
\begin{picture}(56,10)(0,-4)
 \put(0,0){\line(1,0){25}}
 \put(28,0){\circle*{10}}
 \put(31,0){\line(1,0){25}}
\end{picture}
  &=&
 2 \raisebox{+0.1cm}{{\includegraphics{f2pt1tad.eps}}}
+ 2 \; \; \; \raisebox{-0.4cm}{{\includegraphics{f2pt1pm.eps}} }
 + \ldots
\end{eqnarray}
with all other diagrams ignored.

 The next step is  to analyse the
high-$T$ limit of $h(i\omega_n,p)$. For that purpose, we will
consider the DR for the $Q's$ discussed in \cite{grs00}. That is,
we have:
\begin{equation}Q^2 (x,\tau)=\frac{1}{2}\left[\cosh (2\pi T x)-\cos(2\pi T
\tau)\right],
\end{equation}
so that
\begin{eqnarray}
\left[Q^2 (x,\tau)\right]^{-\lambda^2/4\pi}=2^{\lambda^2/2\pi}
e^{-\lambda^2 T \vert x \vert/2}\left[1+e^{-4\pi\vert x \vert T}-2
e^{-2\pi\vert x \vert T}\cos(2\pi T\tau)\right]^{-\lambda^2/4\pi}.
\label{acqsqexp}
\end{eqnarray}

Note that for large distances $\vert x \vert\rightarrow\infty$,
the integrand of (\ref{ach}) is screened with the thermal mass
scale  \bea
 m_T &=& \lambda^2 T/2, \label{mtdef}
\end{eqnarray}
which ensures that the function $h$ is IR finite and in the end
will allow us to obtain IR meaningful results for the
conductivity.

If we expand in (\ref{acqsqexp})

\begin{eqnarray}
[1+A]^{-\lambda^2/4\pi} &=&
 \sum_{k=0}^\infty
 \frac{\Gamma\left[1-\lambda^2/(4\pi)\right]}{\Gamma\left[1-k-\lambda^2/(4\pi)\right] \Gamma(k+1)}
 A^k ,
 \label{acdrexp}
\end{eqnarray}
where $A=e^{-4\pi\vert x \vert T}-2 e^{-2\pi\vert x \vert
T}\cos(2\pi T\tau)$, we get

\begin{eqnarray}
h(i\omega_n,p) &=& \sum_{k=0}^{\infty} \sum_{l=0}^k \sum_{m=0}^l
\delta_{n+l-2m,0} \;  2^{\lambda^2/2\pi}
 \frac{(-1)^l}{(k-l)!(l-m)!m!}
 \nnel
 && \times
 \frac{\Gamma\left[1-\lambda^2/(4\pi)\right]}{\Gamma\left[1-k-\lambda^2/(4\pi)\right]}
 \ \frac{4 \pi(2k-l)+\lambda^2}{p^2 + (2 \pi T (2k-l)+m_T)^2}.
\label{achprev2}
\end{eqnarray}

Our approach will consist in keeping only the dominant terms in
the DR limit ($p\ll 2\pi T$ {\it and} $\lambda^2\ll 4\pi$) for the
sums in (\ref{achprev2}) but keeping the exponential term $\exp
(-m_T \vert x \vert)$ in (\ref{acqsqexp}), in order to reproduce
the correct IR behaviour. For instance, consider the $n=0$
contribution to (\ref{achprev2}). It is not difficult to see that
the dominant contribution is the term $k=l=m=0$ in the above sum,
namely,

\be h(0,p)  \approx  2^{\lambda^2/2\pi}
\frac{\lambda^2}{p^2+m_T^2} .
 \label{achzero}
\ee while the contributions with $k\geq 1$ are subdominant. For
instance, for $k=1$ we have the non-leading contribution

 \beq h^{NLO}(0,p)=
2^{\lambda^2/2\pi}\left(-\frac{\lambda^2}{4\pi}\right)
\frac{\lambda^2+8\pi}{p^2+\left[\left(\frac{\lambda^2+8\pi}{2}\right)
    T \right]^2},
    \end{equation}
which is subdominant compared to (\ref{achzero}) in the limits
$p\ll 2\pi T$, $\lambda^2\ll 4\pi$. In turn, we see why it is
important to take also the small $\lambda$ limit (in the sense
explained above) as we had anticipated in the previous section.
That is, $m_T$ has to be considered as a quantity of the same
order as $p$ and both $m_T,p \ll 2\pi T$. In the language of HTL,
$p$ and $m_T$ are soft scales and $T$ is a hard scale. Thus, we
see why it is consistent to keep $m_T$ without expanding further
in $\lambda$ in the exponential in (\ref{acqsqexp}).

We can do the integral (\ref{ach}) numerically and check our
previous  approximation. For $n=0$, before doing the $x$-integral
in (\ref{ach}) we have in the DR regime

\begin{equation}\int_0^\beta d\tau  \left[Q^2 (x,\tau)\right]^{-\lambda^2/4\pi}
\approx \beta 2^{\lambda^2/2\pi} e^{-\lambda^2 T \vert x \vert/2}.
\label{numx} \end{equation}

\begin{figure}\centerline{\includegraphics{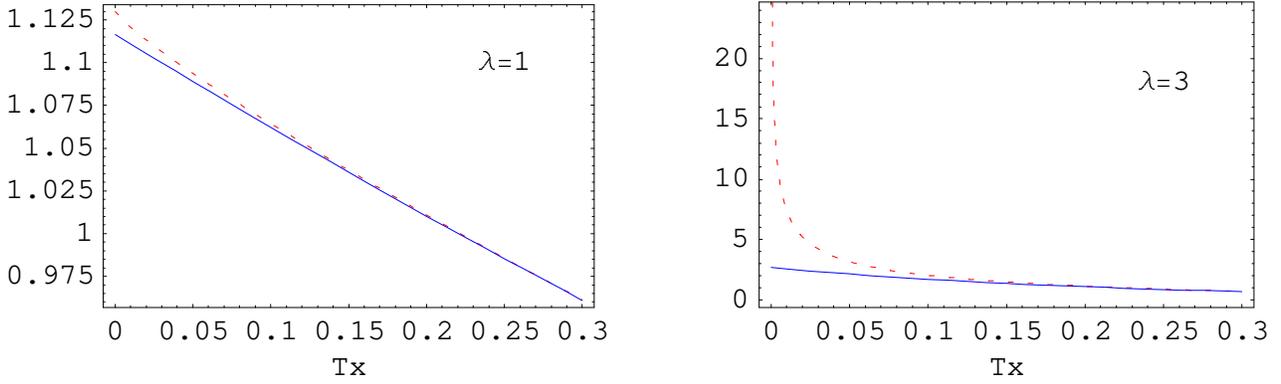}}
\caption{The dashed line is the l.h.s of (\ref{numx}) evaluated
numerically and the solid line is the asymptotic expansion in the
r.h.s, for two different values of $\lambda$.} \label{fignumx}
\end{figure}

The comparison between the l.h.s of (\ref{numx}) numerically
integrated and the asymptotic expression on the r.h.s  is showed
in Figure \ref{fignumx}. The approximation is worse as
$\lambda^2/4\pi$ increases or $Tx$ decreases, as expected.
Nevertheless, for values of $\lambda$ even as large as $\lambda
=3$ it works remarkably well for $\vert x \vert>1/T$. In practice
it is important that the condition $\lambda^2\ll 4\pi$ does not
have to be enforced rigidly, since small $\lambda$ requires large
$g^2$. Although (\ref{realcond}) is valid for all $g^2$, we would
like our result to be more than a super-strong coupling result at
high temperature.


The above procedure can be followed also for the $n\neq 0$ modes,
for which the leading order term in the sum is  $k=l=\vert n
\vert$ (note that $n\neq 0$ implies $k\neq 0$ and hence $l\leq k <
2k$). Expanding also to leading order in $\lambda^2/4\pi$ we get

\begin{eqnarray}
h(i\omega_{n},p)&\approx& 2^{\lambda^2/2\pi} \lambda^2 \left\{
\frac{\delta_{n0}}{p^2+m_T^2}+\sum_{k\neq 0}   \frac{\delta_{nk}}
{p^2+\left[\left(\frac{\lambda^2+4\pi \vert k \vert}{2}\right)
    T\right]^2} \right\}
 \label{hans1a}\\
 &\approx& 2^{\lambda^2/2\pi} \lambda^2 \left\{
\frac{\delta_{n0}}{p^2+m_T^2}+\sum_{k\neq 0}   \frac{\delta_{nk}}
{\omega_k^2} \right\}, \label{hans1b}
\end{eqnarray}
where $\omega_k=2\pi k T$.

An important remark is in order here. The expressions
(\ref{hans1a}) and  (\ref{hans1b}) are {\em equivalent} in the DR
limit. That is, they differ in $\Od(p^2/(2\pi T)^2)$,
$\Od(m_T^2/(2\pi T)^2)$, $\Od(\lambda^2/4\pi)$ terms. However,
that does not mean that their analytic continuations to real
energies will be equally close for all energies. For that reason
we have given them separately because they provide a clear example
of how the analytic continuation of slightly varying
imaginary-time results (including arbitrary non-leading order
terms) can give different physical answers. As a matter of fact,
and for reasons to become clear below,  we will also introduce a
further  modification of \tref{hans1b}:

\be h(i\omega_{n},p) \approx 2^{\lambda^2/2\pi} \lambda^2  \left\{
\frac{\delta_{n0}}{p^2+m_T^2}
\left(\frac{p^2+m_s^2}{m_s^2}\right)^{\lambda^2/4\pi}+\sum_{k\neq
0} \frac{\delta_{nk}}{\omega_k^2}
\left(\frac{\omega_k^2}{m_s^2}\right)^{\lambda^2/4\pi} \right\}
+\Od\left[\frac{\lambda^2}{4\pi}\log
\left(\frac{\lambda^2}{4\pi}\right)\right], \label{hans1c} \ee
where $m_s=\Od(m_T)=\Od(\lambda^2 T)$ is a soft scale mass. We
could take $m_s=m_T$, although \tref{hans1c} allows for more
general situations (see below).

Finally, combining our results in this section with those in
section \ref{sec:low},  we get a high-$T$ expression for the slope
of the MT charge density at the origin simply by replacing in
\tref{exactcd} the coefficient $a_1$ of the $p^2$ term in
$\Sigma_T(0,p)$, which we readily obtain from the zero mode
contribution \tref{achzero}. The expression thus obtained is
consistent with the result found in \cite{grs00} for the charge
density.


\subsection{Analytic continuation to real
energies}\label{sec:ancont}

We now have an expression for the SG imaginary-time propagator
$\Delta_T(i\omega_n,p)$ based on $h(i\omega_n,p)$ of
\tref{hans1a}-\tref{hans1c}. However, all the physical information
we will need for dynamical quantities such as the conductivity is
encoded in the  retarded SG propagator at {\em real} energies. So
one needs to analytically continue from the values
$\Delta_T(i\omega_n,p)$ at discrete imaginary frequencies
$i\omega_n$ to arbitrary complex $\omega$.

The work of Baym and Mermin \cite{BM} showed, as explained in
appendix \ref{appa}, that by enforcing a particular set of
conditions, those given in \tref{BMcond}, one produces a
generalised propagator, $\Delta(\omega_n,p)$, a function of
complex energies $\omega$, whose values at $\omega=E\pm i\epsilon$
with $E$ real, give the retarded and advanced propagators, e.g,
eq.(\ref{relitrp}).

Let us try to apply the procedure of Baym and Mermin to our
expression \tref{sgap1} for the IT propagator, and try to find the
retarded equivalent of \tref{sgap1}.  The generalised propagator
will have the form
\begin{eqnarray}
\Delta(\omega,p) &=& \Delta^0(\omega,p)
 +i\Sigma (\omega,p)\left[\Delta^0(\omega,p) \right]^2.
 \label{gpsd}
\end{eqnarray}
The analytic continuation of the first term in \tref{sgap1}, a
free IT propagator, is well known and used as an example in
Appendix \ref{appa} and in section \ref{fbm}.  The generalised
continued function $\Sigma (\omega,p)$ can either be defined from
\tref{gpsd} as the second term in the full propagator divided
(truncated) by two generalised free propagators, $\Delta^0$.  With
a little work one can see that $\Sigma$ is indeed the analytic
continuation of the IT function $\Sigma_T$, satisfying a slightly
altered set of BM conditions \tref{BMcond} as its large energy
behaviour can be such that $\Sigma\omega^{-4}$ vanishes for large
$\vert \omega \vert$. It is then a short step from this
generalised function to the retarded version using \tref{relitrp},
and we will find the form

\begin{eqnarray}
 \Delta_R(E,p) = \Delta_R^0(E,p)+
  i\Sigma_R (E,p) \left[\Delta_R^0(E,p)\right]^2,
 \label{rpsd}
\end{eqnarray}

where the first term is simply the free retarded propagator \be
\Delta_R^0(E,p) =\frac{i}{(E+i\epsilon)^2 - p^2 - \mu_0^2},
\label{freeprop2} \ee and the second term we can write as a
retarded function multiplied by two free retarded propagators.
Physically, we will be interested only in the small $p$ behaviour
of $\Sigma (\omega,p)$ at high temperatures.

\subsubsection{Approximate analytic continuations}

In the previous sections, we have analysed the IT propagator to
leading order in the high $T$ expansion \tref{acsgap2a}. In fact,
to that order, and according to \tref{sedef}, $\Sigma_T\approx
-\Pi_T$, with $\Sigma_T$ in (\ref{achprev}) and $\Pi_T$  the {\em
imaginary-time} self-energy. One must bear in mind though that the
analytic continuation of perturbative approximations to
\tref{sedef} for arbitrary complex energies must be done
carefully, taking into account the poles of the  free propagator
and the $\Sigma$ functions. We will discuss this issue in section
\ref{quasi} below.

We have been able to find approximate expressions for
$h(i\omega_n,p)$ in the DR limit in \tref{hans1a}-\tref{hans1c}.
However, as  discussed in greater generality in Appendix
\ref{appa}, it is possible to find different functions whose
analytic continuations all give the same IT result approximately,
i.e. up to non-leading order terms in the DR limit. In this
section, we will give explicit examples of such functions for the
case of $h(i\omega_n,p)$. All of them satisfy the conditions
explained in Appendix \ref{appa}, i.e,  such functions
$h(\omega,p)$ are analytic off the real axis, $h\omega^{-4}$
vanish for large $\vert\omega\vert$ and $h(\omega=i\omega_n,p)$
coincides with \tref{hans1a}-\tref{hans1c} {\em to leading order}
in DR. In section \ref{sec:extraphys}, we will show that
additional conditions may be imposed to ensure that the analytic
continuation gives unique and meaningful physical answers. Let us
then consider the following cases:

\begin{enumerate}

\item The first choice we could make is to neglect everything but
the zero mode contribution in \tref{hans1a}-\tref{hans1c}, which
is dominant in DR with respect to the $n\neq 0$ modes, i.e, we
take just $h(i\omega_n,p)= h(0,p) \delta_{n0}$. As discussed in
Appendix \ref{appa} (see \tref{itdr} and \tref{wrongac}), analytic
continuation gives a function $h_1(\omega,p)$ which vanishes
everywhere off the real axis. Therefore, for $E\in\IR$,
\be
h_{1R}(E,p)=h_1(E+i\epsilon)=0, \label{hans1}\ee
 so that only the constant
term $h(0,0)$ would survive in the analytically continued retarded
propagator of (\ref{achprev}). In this regard, we note that when
only the leading zero mode contribution is taken, i.e, when one
replaces just $\left[Q^2(x,\tau)\right]^{-\lambda^2/4\pi}$ by
$2^{\lambda^2/2\pi} e^{-\lambda^2 T \vert x \vert/2}$ in
\tref{acqsqexp}, one can obtain closed expressions for the {\em
full} self-energy sum in \tref{se1}-\tref{itsefac} up to
$\Od(p^2)$ for any $\omega_n$, in terms of the Coulomb gas
pressure and charge density, following similar steps as in
\cite{grs00}. However, as the example of the free propagator in
Appendix \ref{appa} shows, this analytically continued propagator
may be a very bad approximation (we will be more precise below).
More realistic dynamics requires knowledge of the contribution of
the $n\neq 0$ modes as well, even if each heavy mode is of a
smaller order in our approximations than those neglected in
calculating our zero mode contribution.

\item With this in mind, for the  second case we take (\ref{hans1a}) as the
starting point. From this form, one can apply the BM conditions
\tref{BMcond} and derive a unique analytic continuation to real
energies. In this case, as so often, a straight $i\omega_n
\rightarrow \omega$ replacement in the functional form is
essentially sufficient and we find

\begin{eqnarray}
h_2(\omega,p)=
  \frac{2^{\lambda^2/2\pi}\lambda^2}
{p^2+\left[m_T - i s\omega\right]^2},
    && s := \theta\left[\im(\omega)\right] -
    \theta\left[-\im(\omega)\right]
 \label{hans2}
\end{eqnarray}
Note that the because of the modulus on the integer $k$ in
\tref{hans1a}, the form of the generalised function changes in the
upper and lower half planes. This is common and through
\tref{relitrp}, this then corresponds to the retarded and advanced
functions being distinct. Thus,  the retarded function of
\tref{hans2} for real $E$ is $h_{2R}(E,p)=h_2 (E+i\epsilon)$ with
 $s=1$. We remark that the result \tref{hans2} obeys
the BM conditions
 \tref{BMcond}, reproducing all the Matsubara modes in \tref{hans1a} {\em
 including} $n=0$ (see the discussion in Appendix \ref{appa}).

\item If we take  \tref{hans1b} instead of  \tref{hans1a} as our
IT function, then we would have, as our analytic continuation,

\be h_3 (\omega,p)=
\frac{-2^{\lambda^2/2\pi}\lambda^2}{\omega^2-p^2-m_T^2},\label{hans3}
\ee which again gives  $h_3 (\omega,p)\approx h(i\omega_n,p)$  up
to non-leading order terms in DR.

\item Finally, consider the fourth expression, obtained by
analytically continuing \tref{hans1c}:

\be h_4 (\omega,p)=-
\lambda^2\left(\frac{2}{m_s}\right)^{\lambda^2/2\pi}
\frac{\left(p^2+m_s^2-\omega^2\right)^{\lambda^2/4\pi}}
{\omega^2-p^2-m_T^2}.\label{hans4} \ee
\end{enumerate}

Clearly, we could obtain infinite variety in our analytic
continuations just by modifying the original high-$T$ imaginary
time expression by non-leading terms. The important point, as
emphasized in Appendix \ref{appa}, is that the difference in
physical quantities obtained by analytically continuing functions
differing at non-leading order should be also of non-leading
order. The four examples we have shown above are enough to show
that this is not necessarily the case, unless further restrictions
are applied, as we will see below. In fact, note that the analytic
structure of the retarded version of those functions is very
different. For instance,  while $h_2$ in \tref{hans2} for $s=1$
has complex single poles at $\omega=\pm p-im_T$, $h_3$ in
\tref{hans3} has real poles at $\omega=\pm E_p$ with
$E_p=\sqrt{p^2+m_T^2}$ and $h_4$ in \tref{hans4} has  branch cuts
for real $\omega^2>p^2+m_s^2$ in addition to the singular
behaviour at $\omega^2=E_p^2$.
 In general, it is not difficult to realise that by moving the
pole of the AC function in the complex plane by soft amounts, but
preserving the zero mode contribution, the IT values at  $n\neq 0$
remain unchanged to leading order. This can be easily achieved  by
replacing for instance $\left[\omega^2-p^2-m_T^2\right]$ by
$\left[\omega^2-p^2-m_T^2+\omega M_+(\omega,p)\right]$ for $\im
\omega >0$ in the denominator of $h$, with $M_+\ll 2\pi T$ a
regular complex soft function such as the denominator does not
vanish for $\im \omega >0$ and so on for $\im\omega<0$ with
 another function $M_- (\omega)$. Thus, while the
$\omega=i\omega_n$ values are approximately equal, the
contributions near the respective poles are arbitrarily different.
The poles of the retarded propagator are related to physical modes
at $T\neq 0$, or quasi-particles, while the branch cuts of the
self-energy have to do with their decay rate. However, as it is
clear from the previous discussion, we cannot determine the
presence of soft poles unambiguously. There is no novelty in this
as a general problem. For example, the whole programme of Pad\'e
approximants is based on how best to exploit such ambiguities. We
will address this question again in section \ref{quasi} although
in the end the physical results will not be affected by our
ignorance about the precise analytic structure.

\begin{figure}
 \centerline{\includegraphics{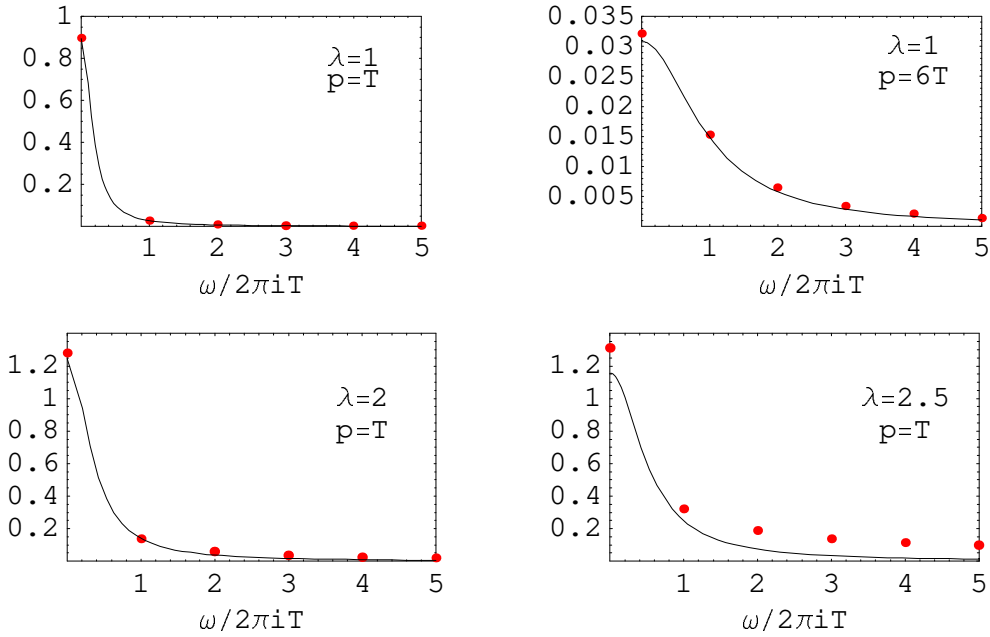}}
 \caption{The solid line
is $h_3(\omega,p)$ as given by \tref{hans3} and the dots represent
the IT values at $\omega=i\omega_n$ calculated numerically from
the definition (\ref{ach}), for $n\leq 5$.} \label{fignumac}
\end{figure}

Before proceeding, let us comment that one can check numerically
that the above functions indeed match the Matsubara modes
approximately. For instance, in Figure \ref{fignumac} we have
plotted the discrete $n$-values against the function
$h_3(\omega,p)$ in \tref{hans3}. We see that the agreement is
quite good, given the numerical uncertainties, for small $p$ and
small $\lambda$, as expected. It gets worse as $\lambda$ or $p$
increase, although it still gives a reasonable agreement for $p
\lesssim 2\pi T$. Note that as $p$ increases, the $n\neq 0$ mode
become of the same importance as the zero mode, but the magnitude
of the latter becomes much smaller. Recall that if we plotted the
approximation \tref{hans1}, it would be just  a function that
vanishes everywhere except at the origin, where it matches the
zero mode. From Figure \ref{fignumac} we see
 that this is a cruder  approximation than $h_3$ in the range
$\omega^2<(2\pi T)^2$. This gives support to the idea, explained
in Appendix \ref{appa}, that we need to combine the information
about the $n\neq 0$ modes, essential for AC, with that on the one
soft mode $n=0$, even if we can only do this approximately.

\subsubsection{A physical condition}
\label{sec:extraphys}

We are interested in the fermion conductivity, which as we have
seen  is related to the SG retarded propagator. In particular, as
we will see in section \ref{phys} (see eq.\tref{condlr4}) the time
evolution of the conductivity is directly related to the Fourier
transform of the retarded propagator:

\be \Delta_R(t,p)=\int_{-\infty}^{\infty} \frac{dE}{2\pi} e^{-i E
t} \Delta_R (E,p), \label{rpft}\ee $\Delta_R$ being given in
\tref{rpsd}, where in the high $T$ limit,

\be \Sigma_R (E,p)\approx \frac{\lambda^2}{2}m^2 T^2
\left(\frac{m^2}{T^2}\right)^{1-\lambda^2/4\pi} \tilde h_R(E,p),
\ee with
$$
\tilde h_R(E,p)= h_R(E,p) - h(0,0),
$$
where $h(0,0) \approx 2^{\lambda^2/2\pi}\lambda^2/m_T^2$ to
leading order.

Consider now the contribution of the second term, i.e the
$\Sigma_R$ term, to \tref{rpft}. As discussed in Appendix
\ref{appa}, the retarded function $\tilde h_R (z,p)$ can be also
analytically continued to complex $z$, simply starting from the
retarded $h_R(E,p)$ for real $E$ and replacing $E$ by $z$. This
function, by construction, has its poles and branch cuts  in the
lower half plane. On the other hand, the generalized analytic
functions $h$ we have been discussing in the previous sections
satisfy Schwartz's reflection principle $h(z^*)=h^*(z)$ and the
property $h(z)=h(-z)$. This means that $h_R (-z,p)=h_R^*(z^*,p)$.

 Therefore, taking $t>0$ and closing the
integration contour from below the real axis, we find the
following contribution to $\Delta_R (t,p)$ given by the poles at
$\omega^2=\omega_p^2\equiv p^2+\mu_0^2$:

\ba \frac{1}{2\pi}\frac{\partial}{\partial \mu_0^2}
\left\{\frac{-2\pi i}{2\omega_p}\left[\tilde h_R(\omega_p,p)
e^{-i\omega_p t}- h_R(-\omega_p,p) e^{i\omega_p
t}\right]\right\}_{\mu_0^2=0} \nonumber\\ =
\frac{\partial}{\partial \mu_0^2}\frac{1}{\omega_p}
\left\{\im\left[\tilde h_R(\omega_p,p) e^{-i\omega_p
t}\right]\right\}_{\mu_0^2=0}\nonumber\\ = \frac{1}{2p^3}\left[
\re \tilde h_R (p,p)\left(\sin pt -p t \cos pt\right) - \im \tilde
h_R (p,p)\left(\cos pt - p t \sin pt\right)\right]\nonumber\\
-\frac{1}{2p^2}\left(\re \tilde h_R' (p,p)\sin pt - \im \tilde
h_R' (p,p)\cos pt\right), \label{tgrow} \ea where $\tilde
h_R'=(\partial/\partial \omega) \tilde h_R (\omega,p)$. Note that
$\tilde h_R$ does not depend on $\mu_0$ (see our comments about
the IR behaviour in section \ref{sec:sgpropit}) and that the
expression \tref{tgrow} is real, as it should, since it
contributes to the conductivity. Therefore, from \tref{tgrow} we
see that $\Delta_R (t,p)$ has a term growing linearly in time (and
so  does the conductivity) unless $\tilde h_R (p,p)=0$ (although
$\tilde h_R' (p,p)$ will be in general  different from zero).

Hence, if we want the conductivity to remain bounded in time, as
seems physically reasonable \footnote{In addition, an unbounded
correction would become eventually larger than the leading order,
so that the high-$T$ expansion would be meaningless.} this gives
us the following condition for the retarded $\Sigma$ function on
the mass shell of the SG massless field:

\be \Sigma_R (E^2=p^2,p)=0 \label{physcond}\ee for
$\epsilon\rightarrow 0^+$.
 This is an
extra, physical, condition, in addition to the BM mathematical
ones discussed in Appendix \ref{appa}.

From the IR behaviour of $h_R$ as $p\rightarrow 0^+$, discussed in
previous sections, it is not difficult to see that once we demand
\tref{physcond}, the rest of the contributions in \tref{tgrow}
remain bounded in time when taking the Fourier transform also in
the $p$ variable. As for the contributions of the poles of $\tilde
h_R$ in the lower half plane, say at $\omega=\pm E(p)-i\gamma (p)$
with $\gamma (p)\geq 0$, their contribution is damped at long time
as $\exp (-\gamma t)$ if $\gamma\neq 0$ (as it is the case for
instance for the retarded version of \tref{hans2} where
$\gamma=m_T$) and one can check that for $\gamma=0$, although
there is no exponential damping, the contributions to $\Delta_R
(t,p)$ remain bounded as well, since one gets $\sin E t$ and $\cos
Et$ instead of $\sin \omega_p t$ and $\cos \omega_p t$ as in
(\ref{tgrow}), so that all the $\mu_0$ dependence is in the
$t$-independent contributions.

Finally, if $h_R (z,p)$ has also a branch cut, as it is the case
for $h_{4R} (z,p)=h_4 (z+i\epsilon,p)$ with $h_4$ in \tref{hans4}
 for $z=-i\epsilon\pm E'$ with $E'>\sqrt{p^2+m_s^2}$, the branch cut
 contribution to $\Delta_R (t,p)$ gives also a bounded
 contribution, as long  as $h_R(E,p)$ is  integrable  for $E\geq
 E_{cut}$ where $E_{cut}$ is the cut endpoint. For instance, for
 the case of $h_{4R}$ with $m_s=m_T$, such a contribution is proportional to

 $$
\int_{E_p}^\infty dE \frac{1}{(E^2-p^2)^2} \frac{\sin (E
t)}{(E^2-E_p^2)^{1-\frac{\lambda^2}{4\pi}}},
 $$
with  $E_p=\sqrt{p^2+m_T^2}$, which is perfectly finite and
bounded in time, since the integrand behaves like $(E-E_p)^{-r}$
with $0<r<1$ near $E_p$. The same is true for $m_s\neq m_T$, as it
is discussed in section \ref{phys}, see eq. \tref{cuteq}.

Of the four cases considered in the previous section, only $h_3$
and $h_4$ satisfy the on-shell condition \tref{physcond} and they
yield respectively:

\ba \Sigma_{3R} (E,p) &\approx& -\mu_T^2
\frac{(E+i\epsilon)^2-p^2}{(E+i\epsilon)^2-p^2-m_T^2}, \nnel
\Sigma_{4R} (E,p) &\approx& -\mu_T^2 \left[\frac{m_T^2
 m_s^{-\lambda^2/2\pi}\left[p^2+m_s^2-(E+i\epsilon)^2\right]^{\lambda^2/4\pi}}{(E+i\epsilon)^2-p^2-m_T^2}
+1\right], \nonumber
 \ea
with

\be\mu_T^2= 2^{1+\lambda^2/2\pi} m^2
\left(\frac{m^2}{T^2}\right)^{1-\lambda^2/4\pi}. \label{mutsq}
\ee

Note that $\mu_T^2\ll m^2$ in DR, but we have not made any
assumption about the relation between $m$ and $m_T$ and therefore
between $\mu_T$ and $m_T$. However, it is natural to take $m$ and
$m_T$ of the same order, since both are soft scales much smaller
than $T$, and therefore $\mu_T^2\ll m_T^2$ . In fact, most of our
results below will be expressed as corrections of order
$\mu_T^2/m_T^2$. This means that the temperatures should be at
least as high as $T\gsim m/\lambda^2$ but they can be even much
higher, which would have simply the effect of making the
perturbative corrections much smaller.

\subsubsection{Analytic structure and quasi-particles}
\label{quasi}

Here, we will discuss  some relevant features of the retarded
propagators we have just obtained. The poles of the retarded
propagator in the lower half plane give the dispersion law for
quasi-particles in the thermal bath \cite{lebellac}, which are the
dual SG bosons in this context. In moving from the MT model, with
two fundamental degrees of freedom, to the sine-Gordon model with
one bosonic degree of freedom, we appear to have lost a degree of
freedom. However, the second bosonic degree of freedom was trivial
and was integrated out when exploiting the duality \cite{gs99}.
For instance, for $m/T \rightarrow 0$, bosonic and fermionic modes
both contribute the same amount to the free energy in 1+1
dimensions. We must take account of this trivial bosonic mode in
the thermodynamic quantities, but for dynamic quantities, only the
single SG mode is necessary for equivalence with the two massive
Thirring modes.

 On the other hand, the branch cuts of the
self-energy $\Pi (E,p)$ along the real axis correspond to the
decay rates of those quasi-particles and the corresponding
discontinuity $\Pi(E+i\epsilon)-\Pi(E-i\epsilon)=2i \im\Pi$ across
the cuts can in principle be calculated by cutting the self-energy
diagrams. Cutting rules and discontinuities in the self-energy at
finite temperature have been extensively analysed  in the
literature \cite{weldon83,kobes86,weldon9802}. To one loop in
perturbation theory \cite{weldon83}  and for the case of a single
particle with mass $m$,  the self-energy has branch cuts at
$s=4m^2,9m^2,\dots$ ($s=E^2-p^2$) corresponding to the decay into
two-particle, three-particle states and so on.  These are the
 $T=0$ cuts  although the discontinuity across the cuts is $T$-dependent.
New "thermal" cuts may appear, even to one loop, corresponding to
processes of emission and absorption of particles in the thermal
bath, not allowed at $T=0$.

In our case,  even to leading order in the high $T$ limit, we are
dealing with a {\em nonperturbative}  resumation of diagrams in
the coupling constant $\lambda$, as for instance in \tref{ach}. As
commented above, this is essential in order to obtain a meaningful
IR behaviour for physical quantities such as the transport
coefficients and it has forced us to use analytic continuations to
approximate imaginary-time results. Thus, we do not have a clear
interpretation in terms of Feynman diagrams that can be cut and
therefore we do not know a priori what should be the analytic
structure of the self-energy on the real axis. To make matters
worse, and as emphasized in previous sections, with our DR
approximation we cannot determine  the position of the self-energy
branch points or the poles of the retarded propagator. One can
move the poles by soft amounts (i.e, of $\Od(m_T)$) and the
analytic continuation will still match the IT  values to leading
order in the DR high-$T$ limit.

The point we want to make here is that, despite this apparent
limitation, we can give meaningful physical predictions. The key
point is that, once the extra condition \tref{physcond} is
imposed, the errors produced by changing the poles or the branch
cuts are perturbatively small. We will check this explicitly in
section \ref{phys}. Before that, let us discuss in some more
detail  the analytic structure of the AC propagators. First, let
us define the retarded self-energy in terms of the retarded
propagator as customary,

\be \Delta_R(\omega,p) = \frac{i}{(\omega+i\epsilon)^2 - p^2 -
\Pi_R(\omega,p)}. \label{defrese}\ee

Note that we cannot write \tref{rpsd} as $\Delta_R\sim
i(\omega^2-p^2+\Sigma_R)^{-1}$ since, as explained in section
\ref{sec:sgpropit},  we do not get an infinite series of $\Sigma$
insertions.  Remember also that the poles of $\Delta_R(\omega,p)$
lie in the lower half of the complex plane $\omega$ and the
retarded propagator for real energies $E$ is $\Delta_R(E,p)$.

Consider first \tref{hans2}, which gives for $\mu_0=0$

\be \Delta_{2R}(\omega,p)=i
\frac{\mu_T^2\left[s+2i(\omega+i\epsilon)m_T\right]+s\left[s-m_T^2+2i(\omega+i\epsilon)m_T\right]}
{s^2 \left[s-m_T^2+2i(\omega+i\epsilon)m_T\right]},\ee where
$s=(\omega+i\epsilon)^2-p^2$ and $\mu_T^2$ defined in
\tref{mutsq}.

We see that $\Delta_{2R}$ has a double pole at $s=0$ and single
poles at $\omega=\pm p - im_T$. On the other hand, the retarded
self-energy defined through \tref{defrese} satisfies $\im \Pi_R
(E+i\epsilon,p)\neq 0$ for any $E$ real different from zero, so
that it has a cut along the whole real axis. Note also that
$\Pi_R(\omega,p)$ is singular for the complex $\omega$ solution of
$\mu_T^2\left[s+2i(\omega+i\epsilon)m_T\right]+
s\left[s-m_T^2+2i(\omega+i\epsilon)m_T\right]=0$.

We compare this with the solution \tref{hans3}, which gives

\be \Delta_{3R}(\omega,p)= i\frac{s+\mu_T^2-m_T^2}{s
\left[s-m_T^2\right]}.\ee

Now, $\Delta_{3R}$ has single (and real) poles at $s=0$ and
$s=m_T^2$ and the self-energy is real on the real axis except at
the singular point $(\omega+i\epsilon)^2=p^2+m_T^2-\mu_T^2$.

The analytic structure of the above two solutions is completely
different. As we have discussed already, the position of the poles
and  branch cuts  change within DR. A different story though is
the physical interpretation of the above results. First, note that
the two solutions above for the retarded propagator have a common
feature: the self-energy is singular close to the new "thermal"
poles (i.e, those different from $s=0$). That is, the distance
between the singular points and the "thermal" poles in the
$\omega$ plane is proportional to $\mu_T$, which is much smaller
than the positions of the poles themselves, which is $\Od(m_T)$.
The values for which the self-energy diverges are related to
pinching or end point singularities \cite{weldon9802} and it is
therefore not clear that those poles should be interpreted as
quasi-particles energies or damping rates. However, the behaviour
of the two above solutions near the pole at $s=0$ is very
different. Thus, while $\Delta_{3R} (s\rightarrow 0)\sim
i(1-\mu_T^2/m_T^2)s^{-1}$ is a massless pole with a $T$-dependent
residue,  $\Delta_{2R} (s\rightarrow 0)\sim i(\pm 2ip
\mu_T^2)/(\pm 2 i p -m_T)s^{-2}$ is a double pole. A double pole
is an indication of the breakdown of the Schwinger-Dyson expansion
around the massless propagator \cite{weldon9802} and in fact  we
have seen in the previous section that it leads to unphysical
behaviour. Note that the condition \tref{physcond}, which is
satisfied by $\Delta_{3R}$ but not by $\Delta_{2R}$, ensures that
the pole at $s=0$ is a single pole. Therefore, our physically
acceptable solution for the retarded propagator is consistent with
a massless dispersion law. However, we want to make clear that our
approach does not allow to say whether the quasi-particles
dispersion law really remains massless or there is a "thermal"
true pole of $\Od(m_T)$ as it happens in the HTL approach. All
that we can say is that $m_T$ plays the role of a screening mass
in the sense explained in previous sections. In addition, note
that the contribution to physical observables from the pole of
$\Delta_{3R}$ at $s=m_T^2$ is reduced by a $(\mu_T^2 /m_T^2)$
factor with respect to that of the $s=0$ pole, so that the thermal
contributions in DR are always perturbatively small with respect
to the leading order. Another indication of the ambiguities in the
interpretation of the retarded propagator is  the result
\tref{hans4}, which also satisfies the condition \tref{physcond}
and  we could have chosen to have a branch cut starting at
$m_s=2m_T$, as expected from a true thermal quasi-particle pole of
mass $m_T$ \cite{weldon9802}.

Summarizing, the DR approach does not allow us to fix uniquely the
position of poles and branch cuts of the retarded propagator.
Although the results suggest the appearance of a thermal mass
scale $m_T$, it is not clear within our approximation scheme
whether this is the mass of the quasi-particles.  Thus we see that
the double-lined "propagator" introduced in section
\ref{sec:sgpropit} really has  a pole but this does not simply
correspond to a stable particle of mass $m_T$. The important point
is that the physics we are interested in, i.e, the conductivity,
does not depend on our choice of functions to be continued
analytically, once the physical condition \tref{physcond} is
imposed. For instance, taking \tref{hans3} or \tref{hans4} will
produce the same answer, to the order we are considering here.


\section{Physical applications. The conductivity.}\label{phys}

Transport coefficients are usually discussed in Fourier space and
the expressions such as \tref{cond1} for conductivity, used to
motivate the calculations of the full  scalar propagator, were in
energy and momentum.  Our interest is in the DC conductivity,
i.e.\ the $\omega,p \rightarrow 0$ limit of \tref{cond1}. However,
the free field propagators $\Delta_0$ appearing in expressions
such as \tref{cond1} are poorly defined in this limit because they
are massless.  The DC conductivity then seems to depend on how the
$\omega,p \rightarrow 0$ limit is approached although, in simple
cases, as in (\ref{realcond}), there may not be any difficulty in
making a choice. For more complicated expressions more care is
needed. Such unsatisfactory behaviour is well known in thermal
field theory for zero-momentum Green functions
\cite{TSEzm,TSEze}, often coming from Landau damping cuts across
zero energy rather than a massless pole as here.

Our solution is to remain in coordinate space where one can
specify a realistic experimental situation, with systems of finite
size examined for finite times.
The lack of analyticity found at zero momentum and energy at
finite temperature can be associated with space-time causality.
After all, $\delta j$ should be zero if we are looking at a point
that is space-like separated from a non-zero electric field. The
usual formulae will then give $\sigma=0$. On the other hand for
the same problem at a time-like separated point, $\sigma$ will be
non-zero. Likewise a simple implementation of a DC conductivity
measurement, the $E,p\rightarrow 0$ limit, implies that a constant
electric field has been applied for all times.  If the system has
finite correlation times, this may be an acceptable approximation
to an experiment where the fields were set up a `long time'
before. However, the presence of poles or cuts, at zero energy and
momentum, suggests that there are long time scales in the problem,
 so one can not simply assume that such a DC current can be
switched on adiabatically in the distant past. In practice,
working in time and space is not especially difficult, and it is
much more physical, so it greatly simplifies the physical
interpretation of real-time calculations in finite temperature
problems.

Turning to our model, we must first specify the electric field.
From Gauss' law the electric field due to a single charge in vacuo
is constant in size, merely switching sign at the location of the
charge, i.e.\ $E \propto q (\theta(x-x_1) - \theta(x_1-x))$ is the
field due to single static charge at $x_1$. A constant field over
a finite region $x_1$ to $x_2$ comes from having two charges of
opposite signs at $x_1$ and $x_2$. Thus we choose \anote{I suggest
to take $t_0=0$ in all this section. It doesn't change the results
and simplifies the discussion}

\begin{equation}\label{Echoose}
  E_{cl}(t,x) = \bar{E} \theta(t-t_0)
  \left[\theta(x-x_1) - \theta(x-x_2)\right],
\end{equation}
where $\bar{E}$ is a constant and $x_2>x_1$ for simplicity, so
that $E=\bar{E}$ if $x_1<x<x_2$, and is zero otherwise.  This is
then a close analogue of the usual large parallel plate capacitor
problem encountered in 3+1 electrostatics.

However, we are turning on the field suddenly, so the two charges
appear instantaneously at time $t_0$. The field is switched on
over a region which is not initially time-like separated, so we
must be careful later with the interpretation. This means that we
have to calculate \ba\label{condlr4}
  \delta j^{(1)}(X) &=&  i  \frac{\lambda^2}{4 \pi^2} \bar{E}
  \int_{t_0}^\infty dt' \int_{x_1}^{x_2} dx' \;
  \int \frac{d^2p}{(2\pi)^2} e^{-ip_0(t-t')} e^{ip(x-x')}
  .(-ip_0) \Delta_R(p_0,p)
  \nnel
  &=&\frac{\lambda^2}{4\pi^2} \bar{E}
   \sum_{j=1,2} s_j \int \frac{d^2p}{(2\pi)^2} e^{-ip_0(t-t_0)} e^{ip(x-x_j)}
    \frac{\Delta_R (p_0,p)}{p},
\ea where $s_1=+1$, $s_2=-1$ and $X\equiv(x,t)$.

As shown in section \ref{sec:dr},  the full retarded propagator is
made up of two parts, as given in \tref{rpsd}. The first part is
the free contribution and for the second one we have obtained
explicit expressions in the DR high-$T$ limit. Let us analyse
those two contributions separately.

\subsection{Free boson term}

Consider the first term, the single free propagator
\tref{freeprop2} and its contribution $\delta j_a^{(1)}(X)$ to the
total $\delta j^{(1)}(X)$.  This gives, for $t\geq t_0$:
\begin{eqnarray}\label{condlr4b}
  \delta j_a^{(1)}(X)
  &=&
   \frac{\lambda^2}{2\pi^2} \bar{E}
  \sum_{j=1,2} s_j \int_0^\infty \frac{dp}{2\pi} \frac{\sin\left[p(x-x_j)\right]}{p}
  \frac{\sin(\omega_p(t-t_0))}{\omega_p}
  \nnel
  &=&
  \frac{\lambda^2}{8 \pi^2} \bar{E}
  \sum_{j=1,2} s_j  \left[ (t-t_0) (\theta_2^j-\theta_3^j) +  (x-x_j)
  \theta_1^j
  \right]   .
  \label{condDani}
\end{eqnarray}
where $\omega_p = \sqrt{p^2+\mu_0^2}$ and in the second line we
have taken the $\mu_0\rightarrow 0^+$ limit after performing the
$p$ integral. This result is clearly IR and UV finite, even if
$\mu_0 \rightarrow 0^+$. However, it is also causal, i.e. the
current is zero if its space-like separated from the electric
field region. The regions where the $\theta$ functions are
non-zero are shown in Figure \ref{f1}. The $\theta_1^j$ is one in
region I - the region inside the forward light cones from the
point $(t_0,x_j)$, i.e. the space-time point where the charge at
position $j$ was switched on. It is zero elsewhere. The
$\theta_2^j$ is then one only in region II to the right (positive
$x$) of this and $\theta_3^j$ is one in region III to the left
(negative $x$). Each of the terms in the sum for $j$ represent the
effects of switching on at $t_0$ the electric field of a single
charge at $x_j$. Note that the regions II and III are space-like
connected to $(t_0,x_j)$ yet the change in the current is non-zero
in these regions.  This is because we have actually switched on
the electric field of these charges instantaneously, and the
static field is non-zero everywhere.  A more realistic experiment
would be to switch on the charges slowly (charging up a parallel
plate capacitor). However this behaviour is not important for the
case at hand. In fact, when we add the two terms in the sum for
$j$ together we get

\begin{eqnarray}
  \delta j^{(1)}_a(X) &=& j_{a,max}
  \left[  \theta_{11} + 2 (t-t_0)L^{-1} \theta_{23} + \left[(t-t_0) -
  (x-x_2)\right]L^{-1}\theta_{21}\right. \nnel
  && \left.   +  \left[(x-x_1) + (t-t_0)\right]L^{-1} \theta_{13}
  \right],
  \label{condlr8}
  \\
  j_{a,max} &=&  \frac{\lambda^2}{8 \pi^2} \bar{E} L,
\label{jamax}
\end{eqnarray}
where
\begin{equation}\label{thetac2}
  \theta_{kl} = \theta_k^1 \theta_l^2, \; \; \; L= x_2-x_1.
\end{equation}
The regions where the $\theta_{kl}$ functions are non-zero are
shown in Figure \ref{f2}.  If the distance between the charges is
$L$, i.e.\ the electric field $\bar{E}$ is on over a region of
length $L$, then the current profile rises until at time
$t-t_0-L/2>0$ it reaches a state where there is a central region
where the current saturates at $j_{a,max}$ and is proportional to
$L$ for constant electric field strength $\bar{E}$.  This region
extend a length $t-t_0-L/2$ either side of the mid-point of the
electric field region for $t>L$.  The current profile then drops
linearly to zero at a distance $t-t_0+L/2$ either side of the
mid-point. Its then zero beyond this, as it must as these regions
are not in causal contact with the region of electric field (i.e,
the regions $22$ and $33$ in Figure \ref{f2}). This is shown in
Figure \ref{f3}.

Thus there are two conclusions.  After an initial rise time, the
current in the region of electric field is constant proportional
to $L$.  Thus the contribution to the conductivity from the first
term in the full propagator is
\begin{equation}\label{sigma}
  \sigma = \frac{\lambda^2L}{8 \pi^2}.
\end{equation}
as in (\ref{realcond}).  Thus the current is always proportional
to the voltage, $\bar{E}L$, and independent of $L$ so the
conductance (the inverse of the resistance) is constant, at
$G=\lambda^2/8\pi^2$, as in (\ref{G}), whatever the length of the
material is studied! However, life is not quite so simple. Note
that in fact the current is also nonzero outside the region of the
electric field. In fact, the current is reaching the same constant
level everywhere as fast as causality allows. This is nothing but
the effect of the SG massless effective mode, propagating at the
speed of light, since $dx/dt=1$ for the extreme points in Figure
\ref{f3}.  Note also that the result
 \tref{sigma} is consistent with our analysis in section \ref{fbm},
 eq.\tref{realcond}. Let us study now the behaviour of the
 temperature corrections to this result in the DR, high-$T$ limit.

\setlength{\unitlength}{1pt}
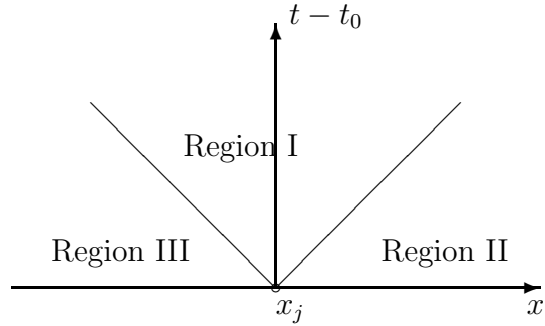
\begin{figure}[htb]
\begin{center}
\begin{picture}(220,110)(-110,-10) \thicklines
\put(-100,0){\vector(1,0){200}} \put(0,0){\vector(0,1){100}}
\thinlines \put(0,0){\line(1,1){70}} \put(0,0){\line(-1,1){70}}
\put(0,0){\circle{3}} \put(-35,50){Region I}
 \put(40,10){Region II}
  \put(-85,10){Region III}
 \put(0,-10){$x_j$}
  \put(5,100){$t-t_0$}
   \put(95,-10){$x$}
\end{picture}
\end{center}
\caption{Light cone regions for single charge}\label{f1}
\end{figure}

\setlength{\unitlength}{1pt}
\begin{figure}[htb]
\begin{center}
\begin{picture}(220,110)(-110,-10) \thicklines
\put(-100,0){\vector(1,0){200}} \put(0,0){\vector(0,1){100}}
\thinlines \put(-25,0){\line(1,1){70}}
\put(-25,0){\line(-1,1){70}} \put(+25,0){\line(1,1){70}}
\put(+25,0){\line(-1,1){70}}
 \put(-30,60){Region 11}
 \put(-7,5){23}
 \put(-60,50){13}
  \put(+40,50){21}
   \put(50,10){Region 22}
    \put(-110,10){Region 33}
 \put(-30,-10){$x_1$}
 \put(-25,0){\circle{3}}
  \put(+30,-10){$x_2$}
\put(+25,0){\circle{3}} \put(0,-3){\vector(-1,0){25}}
\put(0,-3){\vector(1,0){25}}
 \put(0,-12){$L$}
  \put(5,100){$t-t_0$}
\put(95,-10){$x$}
\end{picture}
\end{center}
\caption{Light cone regions for two charges}\label{f2}
\end{figure}
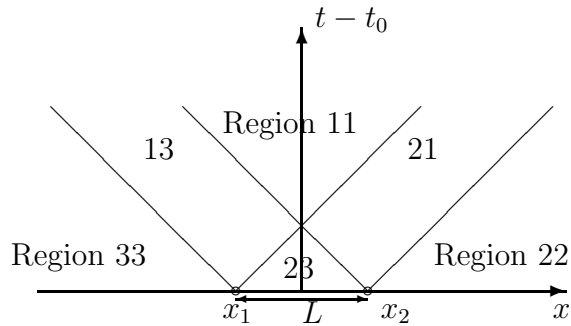

\setlength{\unitlength}{2pt}
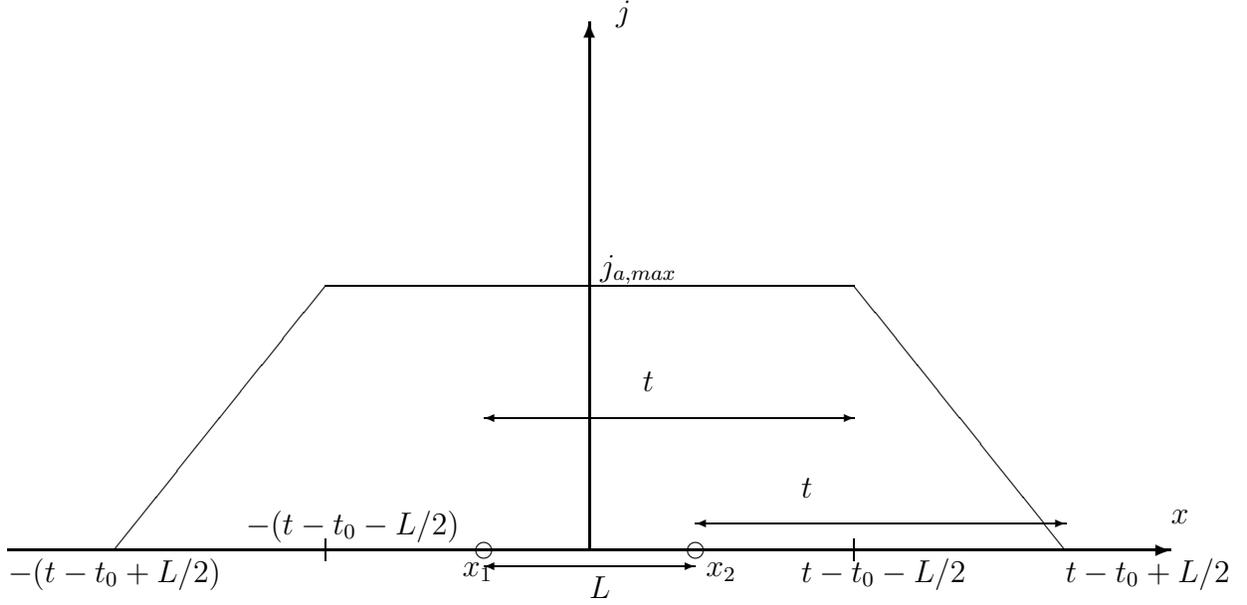
\begin{figure}[htb]
\begin{picture}(220,110)(-110,-10)
\thicklines \put(-110,0){\vector(1,0){220}}
\put(0,0){\vector(0,1){100}} \thinlines
\put(-50,50){\line(-4,-5){40}} \put(50,50){\line(4,-5){40}}
\put(-50,50){\line(1,0){100}} \put(50,-2){\line(0,1){4}}
\put(-50,-2){\line(0,1){4}} \put(40,-6){$t-t_0-L/2$}
\put(-65,3){$-(t-t_0-L/2)$} \put(90,-6){$t-t_0+L/2$}
\put(-110,-6){$-(t-t_0+L/2)$} \put(2,52){$j_{a,max}$}
\put(-24,-5){$x_1$} \put(-20,0){\circle{3}} \put(+22,-5){$x_2$}
\put(+20,0){\circle{3}} \put(0,-3){\vector(-1,0){20}}
\put(0,-3){\vector(1,0){20}} \put(0,-9){$L$} \put(5,100){$j$}
\put(110,5){$x$} \put(+55,5){\vector(1,0){35}}
\put(+55,5){\vector(-1,0){35}} \put(+40,10){$t$}
\put(+15,25){\vector(1,0){35}} \put(+15,25){\vector(-1,0){35}}
\put(10,30){$t$}
\end{picture}
\caption{Current profile (first, free term), $t>t_0+L$}\label{f3}
\end{figure}

\subsection{High-$T$ corrections}

In section \ref{sec:dr} we have discussed the DR high-$T$ (and
large distances) approximation to the retarded propagator. Let us
write the contribution of the second term in the r.h.s of
\tref{rpsd} to the current in \tref{condlr4} as

\begin{eqnarray}
\delta^{(1)} j_b (X)
 &=&-
  i \bar{E}
  \frac{\lambda^2}{4\pi^2}\mu_T^2
  \frac{\partial \Xi(X)}{\partial \mu_0^2},
 \label{currentsec}
\end{eqnarray}
with $\mu_T^2$ given in \tref{mutsq}. Only the cases $h_2$ and
$h_4$ of (\ref{hans2}) and (\ref{hans4}) satisfy the physical
constraint (\ref{hans1}). Taking them in turn, $h_2$ of
(\ref{hans2}) gives
\begin{eqnarray}
\Xi (X)&=&\sum_{j=1,2} s_j \int \frac{d^2p}{(2\pi)^2}
e^{-ip_0(t-t_0)} e^{ip(x-x_j)}
    \frac{1}{p} \frac{1}{(p_0+i\epsilon)^2 - p^2-\mu_0^2}
    \nnel
    &\times&
\left[\frac{m_T^2}{p^2+m_T^2-(p_0+i\epsilon)^2} - 1
 \right],
\end{eqnarray}

\begin{figure}
 \centerline{\includegraphics{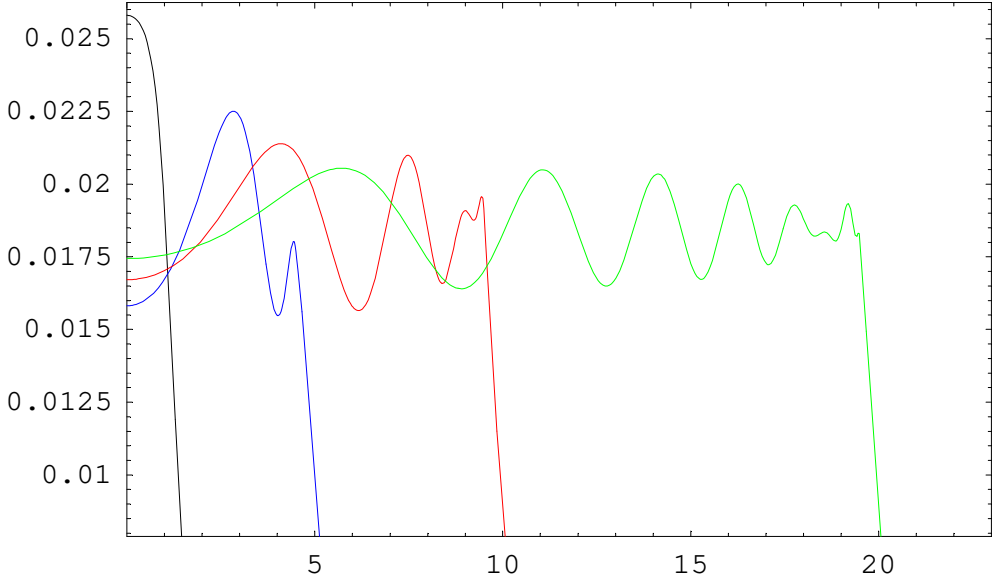}}
 \caption{Time evolution of
the second term $-\delta^{(1)} j_b (X)/j_{max}$. The horizontal
axis is $mx$ and $T=5m$, $L=1/m$ and $\lambda=1$.  The curves in
each plot are for times $tm=1.5, 5, 10$ and $20$ with the later
time plots extending further in $mx$. }
 \label{fjb2}
\end{figure}

When  the $p_0$ integral in \tref{currentsec} is evaluated by the
Residue Theorem for $t\geq t_0$, there is a contribution coming
from the poles at $p_0=\pm\omega_p-i\epsilon$ and another one from
the poles at $p_0=-i\epsilon \pm \sqrt{p^2+m_T^2}$. The second one
is always bounded. The remaining $p$-integrals can be expressed in
terms of Bessel functions as

\ba \delta^{(1)} j_b (X)&=&-\frac{\mu_T^2}{m_T^2} \left\{\delta
j^{(1)}_a(X) \right.\nnel &-& \frac{\lambda^2}{8\pi^2} \bar{E}
\sum_{j=1,2} s_j \sgn (x-x_j)
 \left[\theta(t-\vert x-x_j
  \vert)\int_0^{\vert x-x_j \vert} du
J_0 \left[m_T\sqrt{t^2-u^2}\right]\right.\nnel &+&
\left.\left.\theta(\vert x-x_j \vert - t)
 \int_0^{t} du J_0 \left[m_T\sqrt{t^2-u^2}\right] \right]\right\}.
\label{theta0}
  \ea
for $t_0=0$. The first term in the r.h.s of \tref{theta0}
renormalises the maximum value \tref{jamax} for the free current.
This is the effect of the renormalization of the residue of the
free propagator discussed in section \ref{quasi}. As commented
above, this gives rise to a current that in time-like regions
rises to a constant independent of the length, i.e.\ a constant
resistance. The second term in \tref{theta0} is the effect of the
"massive" mode $m_T$, yielding a causal oscillatory behaviour that
is bounded in time. Our results are shown in Figure \ref{fjb2}.

Finally, let us discuss the result of the calculation using
\tref{hans4} for the analytic continuation of the propagator,
which also satisfies the physical condition \tref{physcond}.
Taking $m_s>m_T$ for definiteness, it is not difficult to see that
the corresponding $\delta^{(1)} j_b$ has three contributions. The
first one is the pole at $p_0^2=\omega_p^2$ which gives again a
renormalization of the factor multiplying $\delta^{(1)} j_a$ which
becomes $1-\mu_T^2/m_T^2+(\lambda^2/4\pi)\mu_T^2/m_s^2$ and
therefore is of the same order as the first contribution in the
r.h.s. of \tref{theta0} (remember that $m_s$ and $m_T$ are of the
same order and $\mu_T^2\ll m_T^2$) ). The second contribution
comes from the poles at $p_0^2=p^2+m_T^2$ and gives exactly the
same as the second and third  terms in the r.h.s. of
\tref{theta0}, multiplied by the factor
$(1-m_T^2/m_s^2)^{\lambda^2/4\pi}$. Finally, the third
contribution is given by the cut for $p_0^2>p^2+m_s^2$ and gives

\ba \left[\delta^{(1)} j_b\right]_{cut}&=&{\bar E}
\frac{\lambda^2}{\pi^3}\sin(\lambda^2/4)\frac{\mu_T^2
m_T^2}{m_s^5}\sum_{j=1,2} s_j \int_0^\infty dp \frac{\sin
p(x'-x'_j)}{p}\nnel&\times&\int_{\sqrt{p^2+1}}^\infty dE
\frac{\left(E^2-p^2-1\right)^{\lambda^2/4\pi}\sin E
t'}{(E^2-p^2)^2 \left(p^2-E^2+m_T^2/m_s^2\right)},\label{cuteq}\ea
for $t_0=0$, where $t'=m_s t$ and $x'=m_s x$. This latter
contribution, because of the $\sin(\lambda^2/4)$ factor,  is
negligible in the small $\lambda$ limit compared to the other two.
We have checked the equivalence between the results obtained with
these two representations of the propagator, numerically for
several values of the parameters in the DR limit. Therefore, we
confirm explicitly that the difference between the physical
transport coefficients calculated with different versions of the
analytically continued propagator is negligible within the DR
limit, as long as the condition \tref{physcond} is enforced.

The end result is that, once the transients have passed,
\begin{eqnarray}
  \delta j^{(1)}(X) &=& j_{max}
  \left[  \theta_{11} + 2 (t-t_0)L^{-1} \theta_{23} + \left[(t-t_0) -
  (x-x_2)\right]L^{-1}\theta_{21}\right. \nnel
  &&\left.   +  \left[(x-x_1) + (t-t_0)\right]L^{-1} \theta_{13}
  \right]
  \label{condlr9}
\end{eqnarray}
where
\begin{equation}
  j_{max} \approx  \frac{\lambda^2}{8 \pi^2}\bigg(1-\frac{\mu^2_T}{m^2_T}\bigg) {\bar E} L
\label{jamax2}
\end{equation}
with $\theta_{kl} = \theta_k^1 \theta_l^2$ defined as before. For
high enough temperature, $\mu_T^2\ll m_T^2$, and so the effect of
the non-leading terms in the temperature expansion is relatively
small.  This describes the regime in which chiral symmetry is
approximately restored (the `molecular phase' of
\cite{grs00,yoshida1,yoshida2}) and for that reason gives results
close to those of the massless Thirring model. This condition is
satisfied in all our plots. At fixed temperature the effect
increases as the fermionic coupling becomes strong.

 In terms of the dual boson parameters
the conductance $G$ of the fermion field is now
\begin{equation}
  G \approx  \frac{\lambda^2}{8
  \pi^2}\bigg(1-\frac{\mu^2_T}{m^2_T}\bigg)\approx\frac{\lambda^2}{8
  \pi^2}\bigg(1-\frac{2^{3+\lambda^2/2\pi}}{\lambda^4}\bigg(\frac{m^2}{T^2}\bigg)^{2-\lambda^2/4\pi}\bigg)
\label{G2}
\end{equation}
However, in terms of the fermionic mass $m$ and coupling constant
$g^2$, as would follow from a diagrammatic description in terms of
fermion loops, we find high temperature corrections to $G$ of
(\ref{G}) of the form
\begin{equation}
G \approx \frac{1}{2\pi(1+g^2/\pi)}\bigg[1-2^{(5\pi
+3g^2)/(\pi+g^2)}\frac{(1+g^2/\pi)^2}{16\pi^2}\bigg(\frac{m^2}{T^2}\bigg)^{(\pi+2g^2)/(\pi+g^2)}\bigg].
\end{equation}
The next nonleading term is down by powers  of $m/T$.

The power of duality in resumming series in $g^2$ that would be
unobtainable otherwise is very striking, and is our main practical
result.
\section{Conclusions}

We have analyzed the fermion conductivity at finite temperature in
the massive Thirring/sine-Gordon (MT/SG) models. We have shown
that the use of the dual degrees of freedom can avoid the usual
problems associated with the infra-red behaviour of transport
coefficients at finite temperature. In this case, the SG boson is
the relevant quasi-particle mode in the limit of strong fermion
coupling constant and  the fermion conductivity can be related to
the boson retarded propagator, which we have analyzed in detail.

First, we have studied the imaginary-time SG boson propagator. Its
leading order for small momentum at zero frequency has been
related to the MT fermion charge density. Particular attention has
been paid to the high $T$ limit, for small boson coupling
$\lambda$ (dimensional reduction). In this limit, a thermal scale
$m_T=\lambda^2 T/2$ emerges. This is the scale which screens the
large distance behaviour of the propagator and therefore ensures
that the results for the transport coefficients are well behaved
in the IR limit. This is a common feature of high-$T$ expansions,
where the $m_T$ scale comes from  high-$T$ diagrams of arbitrarily
higher order in the coupling constant. In this sense, our results
provide a partial resummation in $\lambda$ which ensure the IR
finiteness.

We have paid particular attention to the problem of the analytic
continuation of the approximate (in DR) imaginary-time results.
 We have shown that by imposing the standard mathematical
 conditions to the propagator only, one cannot guarantee that the
 difference between  analytically continued propagators of approximately
 equal imaginary-time propagators remains small. However, one can ensure that physical
 results remain insensitive to this choice by imposing extra
 physical conditions. In particular, we have shown this in detail
 for the conductivity, where we demand
 that it remains bounded in time. We have also discussed the
 role of the "thermal" modes in the retarded propagator. Because
 of the ambiguities associated with the analytic continuation of approximate results,
 the position of poles and branch cuts in the retarded propagator
 cannot be determined with this approach. We have considered specific
 examples, where the analytic structure differs but the
 imaginary-time values are equivalent to leading order in DR. Our
 results are not in contradiction with considering $m_T$ a true
 thermal mass, but are also compatible with massless
 quasi-particles, since $m_T$ is a soft quantity.
 The crucial point is that none of these interpretations prohibits us
  obtaining physically meaningful results for the
 conductivity. However, on the formal side, it is
disappointing that we have had to apply to many approximations in
what is an ``exactly solvable'' model.

 Once the physical condition is imposed,
 the fermion conductivity for this model can be studied
 perturbatively at high $T$. We have found that the resistance
 remains approximately constant for long times, inside the causality
 region. The free SG field contribution correspond to an exactly
 constant resistance, while
 the high $T$ corrections renormalize that constant  and
 generate also bounded transient oscillations.

Thus, we have managed to study in detail several aspects of this
problem which will be generic to more realistic problems where the
imaginary-time formalism is used. In particular, we have
highlighted the importance of the analytic continuation of {\em
approximate} results and the IR zero Matsubara mode. Studies of
analytic continuation for propagators \cite{BM} or higher-order
Green functions \cite{TSEnpt} are usually made in terms of exact
functional forms and do not depend on the values at zero Matsubara
energies.  As we have noted, the IR sector is of vital importance
to the physics and the only direct piece of information in this
sector is in fact the zero Matsubara mode, the one not used in
determining the analytic continuation. We have therefore proposed
an additional condition to be used in these cases, \tref{EGRScond}
which combined to physical conditions such as \tref{physcond}
allows us to find unambiguous physical predictions.


\section*{Acknowledgements}

RR and TSE thank PPARC for financial support.  RR, TSE and DAS
thank the Universidad Complutense of Madrid for hospitality and
financial support, the ESF for support through its COSLAB
programme, and the Rockefeller Foundation at Bellagio for
hospitality, where this work was completed. TSE is grateful to
CERN for a Visiting Fellowship during which part of this work was
done. DAS is grateful to the University of Geneva where part of
this work was also done.  All the authors thank the University of
Salerno, in particular through the ERASMUS/SOCRATES programme, for
hosting some of our discussions. AGN thanks financial support from
the Spanish CICYT project FPA2000-0956.






\appendix

\section{Thermal propagators and Analytic Continuation}\label{appa}

The standard analysis is given in several places, such as
 \cite{lebellac}.  We repeat it here to fix our notation and
definitions, but also because we wish to go beyond the usual
discussions to see how best to deal with approximate results.

Take $t \in \bbC$, complex and lying on a given directed contour
$C$ with ends separated by $-i \beta$.  Define the thermal
propagator as
\beq
 \Delta_C (x,t)\equiv
 \langle \langle T_C \phi (x) \phi (0) \rangle \rangle
 \equiv \Delta^> (x,t) \theta_C (t)+\Delta^< (x,t) \theta_C (-t),
\end{equation}
where  $\Delta^> (x,t)= \langle \langle \phi (x) \phi (0) \rangle
\rangle$, $\Delta^< (x,t)= \langle \langle \phi (0) \phi (x)
\rangle \rangle$, and $\theta_C$ means time ordering along $C$ as
usual. The KMS conditions are then \beq \Delta^> (x,t)=\Delta^<
(x,t+i\beta).
\end{equation}
The Fourier transforms of the propagators are defined to be \beq
\Delta^> (x,t)=\int_{-\infty}^{+\infty} \frac{d\omega}{2\pi}
\int_{-\infty}^{+\infty} \frac{d p}{2\pi} e^{-i\omega t} e^{i px}
 \Delta^> (\omega,p),
\end{equation}
and similarly for other propagators. Then, KMS conditions in
momentum space read \beq \Delta^< (\omega,p)=e^{-\beta \omega}
\Delta^> (\omega,p)
\end{equation}

Now we introduce the spectral function
\begin{equation}
\rho (\omega,p)=\Delta^> (\omega,p)-\Delta^< (\omega,p)=
\left(1-e^{-\beta\omega}\right)\Delta^> (\omega,p). \label{sfdef}
\end{equation}
Note that in position space $\rho(x,t)=\langle \langle
[\phi(x),\phi(0)]\rangle \rangle$. On the other hand, $\Delta^>
(x,t)=\Delta^< (x,-t)$ so that $\rho (x,-t)=-\rho(x,t)$ and then
$\rho(\omega,p)=-\rho(-\omega,p)$ so that
\begin{equation}
\int_{-\infty}^{\infty} \frac{d\omega'}{2\pi} \rho (\omega',p)=0.
\label{sfprop}
\end{equation}

All versions of the propagator can be obtained from the spectral
function. For instance, from (\ref{sfdef}) one readily has
\begin{eqnarray}
\Delta^> (\omega,p)&=&\rho(\omega,p)
\left[1+f(\omega)\right],\nonumber\\ \Delta^<
(\omega,p)&=&\rho(\omega,p)f(\omega), \label{rhobose}
\end{eqnarray}
where
\begin{equation}f(x)=\frac{1}{e^{\beta x}-1}\label{bose}
\end{equation}
is the Bose-Einstein distribution. The simplest example is that of
a {\it free} scalar field of mass $m$, for which
\begin{equation}\rho_0 (\omega,p)=2\pi\epsilon(\omega)\delta
(\omega^2-p^2-m^2).
\label{freerho}
\end{equation}

More generally, let us introduce now the retarded and advanced
propagators
\begin{eqnarray}
\Delta_R (x-x',t-t') &=& \theta(t-t')\langle \langle
[\phi(x,t),\phi(x',t')]\rangle \rangle
\label{apdef2}
\\
 \Delta_A (x-x',t-t') &=& \theta(t'-t)\langle \langle
[\phi(x',t'),\phi(x,t)]\rangle \rangle
\end{eqnarray}
which we define for {\it real} time $t-t'$ only because of the
non-analytic nature of the Heaviside functions.

Using the representation for the step function \beq \theta
(t)=i\int_{-\infty}^{\infty} \frac{d\omega'}{2\pi}
\frac{e^{-i\omega' t}}{\omega'+i\epsilon}
\end{equation}
with
$\epsilon\rightarrow 0^+$, the retarded propagator can also be
written in terms of the spectral function as
\begin{equation}
\Delta_R (\omega,p)=i\int_{-\infty}^{\infty} \frac{d\omega'}{2\pi}
\frac{\rho(\omega',p)}{\omega-\omega'+i\epsilon}. \label{retaux1}
\end{equation}

Now unlike the retarded function in real-time with its Heaviside
functions, this energy-dependent retarded function has a simple
extension into the complex energy plane. It is convenient to
define a general function of complex energy
\begin{equation}
 \Delta (z,p) =
 - \int_{-\infty}^{\infty} \frac{d\omega'}{2\pi}
 \frac{\rho(\omega',p)}{z-\omega'}.
 \label{Deltadef1}
\end{equation}
This seems a little trivial as it simply related to both retarded
and advanced functions through
\begin{eqnarray}
 \Delta_R (\omega,p) &=& -i\Delta (\omega+i\epsilon,p)
 \nonumber\\
 \Delta_A (\omega,p) &=&  i\Delta (\omega-i\epsilon,p) .
 \label{relitrp}
 \end{eqnarray}
 with $\omega\in\IR$.
The point is that, on assuming that the grand canonical average
ensures uniform convergence \cite{TSEnpt} of the thermal traces,
we are assured that this generalised propagator function $\Delta
(z,p)$ is bounded and analytic for all complex energies $z$ except
for real $z$, and $\Delta$ must tend to zero as $|z| \rightarrow
\infty$ \cite{BM,TSEnpt}.  These are properties we will exploit
below so $\Delta$ is a useful intermediate object to work with.

Finally, let us introduce the {\it imaginary time} propagator
$\Delta_T\equiv \Delta_C$ when $C\equiv[0,-i\beta]$. Then,
changing variables to $\tau=it\in [0,\beta] \subset \bbR$ gives
\begin{equation}
 \Delta_T (x,\tau) =
 \Delta^> (x,-i\tau)=
 \int_{-\infty}^{+\infty}
 \frac{d\omega}{2\pi} \int_{-\infty}^{+\infty} \frac{d p}{2\pi}
 e^{-\omega \tau} e^{i px} \rho(\omega,p)(1+f(\omega)).
 \label{dtaux1}
\end{equation}
Therefore, the Fourier transform of the IT propagator \beq
\Delta_T (i\omega_n,p)=\int_0^\beta d\tau \int_{-\infty}^{\infty}
dx e^{-i px} e^{i\omega_n \tau} \Delta_T (x,\tau)
\end{equation}
with $\omega_n=2\pi n/\beta$, reads, from (\ref{dtaux1}),
\begin{equation}
 \Delta_T (i\omega_n,p)=
 -\int_{-\infty}^{\infty}
 \frac{d\omega'}{2\pi} \frac{\rho
 (\omega',p)}{i\omega_n-\omega'}.
 \label{itspectral}
\end{equation}
It is crucial that we remember that in energy coordinates, the IT
propagator is a sequence of functions of $p$ and not actually a
differentiable function of a continuous energy variable.  This
subtlety is easily missed since it is clear that we can relate the
generalised propagator function, $\Delta$, to the IT  propagator
through
\begin{eqnarray}
 \Delta (z=i \omega_n,p) &=& \Delta_T (i\omega_n,p), \;
 \; \; n \in \bbZ
 \label{relitgp}
 \end{eqnarray}

The key to analytic continuation is to note that we can not do the
reverse easily, i.e.\ from the IT propagator we can not use this
relationship alone to determine the full generalised propagator
$\Delta$ at all complex energies.  This is particularly confusing
as analytic calculations, such as here, never give us a sequence
of functions of $p$ for a calculation of the IT  propagator.
Instead, we write down a function of a continuous complex
variable, $\omega_n$, and then note that for the IT propagator it
is only to be taken at Matsubara energies.  The obvious analytic
continuation is to drop the restriction to Matsubara frequencies
in such IT propagator expressions, and inspired by \tref{relitgp}
we would guess that we have then found the generalised propagator
$\Delta$. Unfortunately, this analytic continuation procedure is
not unique as it stands. For instance, multiplying the IT
propagator by arbitrary factors of $\exp(i\beta\omega_n)=1$ would
give different results for $\Delta$ but would not alter the values
at Matsubara frequencies, so that \tref{relitgp} holds.

There is, however, a way forward and we can find a scheme for
analytically continuing from the IT function that gives the unique
function $\Delta$.  The necessary and sufficient conditions were
first stated by Baym and Mermin \cite{BM}. We simply quote here
the result: given the IT discrete propagator $\Delta_T
(i\omega_n)$, then the {\it unique} function $\Delta (\omega)$
satisfies the following BM conditions:
\begin{eqnarray}
1. && \Delta(z=i\omega_n)=\Delta_T (i\omega_n),
 \; \; \; |n| \in \bbZ^+
\nnel  2. && \lim_{|z| = \infty} \left\{ \Delta (z,p) \right\} = 0
\; \; \; \forall \; \im (z) \neq 0 \nnel  3. && \Delta(z,p) \mbox{
is analytic } \; \; \forall \; \im (z) \neq 0, \pi
 \label{BMcond}
\end{eqnarray}


Therefore, if the result of an imaginary-time calculation can be
written as $\Delta_T(z,p)$, analytic off the real axis and
satisfying the second BM condition, it is then guaranteed to be
the one generalised propagator function $\Delta(z,p)$.

The retarded and advanced functions for real energy are then given
by \tref{relitrp}, and a simple Fourier transform will then give
the real-time retarded and advanced functions required for
dynamical problems \cite{lebellac}. This is the usual situation
when dealing with {\em perturbative} calculations, when the
Matsubara sums in the loops are performed first \cite{lebellac}.
However, the key point here is that we are dealing with {\em
nonperturbative} expressions, like (\ref{ach}), which are
nonperturbative in $\lambda$, so that this procedure is not valid
and finding the appropriate analytic continuation of the original
imaginary-time expression is not an easy task.

For instance, consider the free field case. The spectral function
$\rho_0$ of (\ref{freerho}), when inserted in (\ref{itspectral}),
gives the sequence of functions
\begin{equation}
\Delta_T^0 (i\omega_n,p)=\frac{1}{\omega_n^2+p^2+m^2} , \; \; \; n
\in \bbZ\label{acfreeit}
\end{equation}
One quickly sees that the function
\begin{equation}
\Delta^0 (z,p)=\frac{-1}{z^2-p^2-m^2} \label{freerp2}
\end{equation}
obeys all the BM conditions \tref{BMcond} and therefore is the
unique generalised propagator.  The retarded propagator then
follows from \tref{relitrp} as
\begin{equation}
\Delta_R^0 (\omega,p)=\frac{i}{(\omega+i\epsilon)^2-p^2-m^2}.
\label{freerp}
\end{equation}
This is temperature independent, as it should since the dispersion
law for free particles does not depend on the medium properties.

\subsection{Analytic continuation of approximate results}

The results quoted above are all very well for a well defined
function, such as the free propagator. However, in quantum field
theory one does not have exact results for the Imaginary-Time
Formalism (ITF) propagator at Matsubara frequencies.  If we had
been doing a numerical Monte Carlo calculation, the errors are
partly statistical and random and partly come from the fact that
only a finite number of Matsubara frequencies are calculated
directly. This severely limits the accuracy of any possible
analytic continuation from numerical data.

In an analytic calculation the errors are usually of a functional
form, that is we assume that the approximate result differs from
the true result by an amount given by a suitable analytic
function. Indeed, we rarely calculate a sequence of numbers, but
we invariably find a meromorphic functional form, even though we
know it to be approximate and that it is strictly only valid for
Matsubara frequencies.  By checking that the function obeys the BM
conditions \tref{BMcond},
we are then guaranteed that we have already found the relevant AC
of our approximate IT propagator to the whole complex plane and
hence the approximate retarded propagator.

However, there is an interesting question one can ask.  Suppose
the functional error is order $\eta$ where $\eta$ may be a small
coupling, a ratio of a mass scale to the high temperature, or
whatever.  Do the higher order, $O(\eta)$, corrections at the
Matsubara frequencies also lead to small corrections to its
analytic continuation for all complex energies?  In our case are
the functional corrections at Matsubara frequencies, when
analytically continued, still small at real energies?

In fact the answer to this, posed in this simple minded way, is
no. A small correction to the effective mass (real or complex) is
a genuinely small change to the value of a propagator at all
Matsubara frequencies.  However, at real energies, near
mass-shell, the propagator with and without the small shift will
differ by large amounts. See our discussion in sections
\ref{sec:dr} and \ref{phys}, or even the case of the free
propagator analyzed below.

Luckily though, this example shows that we are not interested in
reproducing the analytically continued form of the function such
that the errors in the value of the function are always bounded.
Our goal is that the physics obtained from such functions be well
represented. Our functions are Green functions that are not
directly physical, but  tools used to calculate scattering rates
etc.\ It is not important if our Green function differs by an
infinite amount from the true function at some real energies,
provided the physical predictions are accurate. For example, our
experience tells us that any error in our knowledge of the exact
value of a mass leads only to comparatively small errors in cross
sections. What mattered to the physics was that the propagator
always had a single pole that is important and a small shift in
its position is going to be a small error in physical results. In
our case, the conductivity will turn out to pick up small
$O(\eta)$ errors even if these come from shifts in the mass value
which leads to large differences in the propagators at certain
values (near mass shell).

We have highlighted this example as this is the task we face in
our model.  Namely we calculate an exact ITF propagator but to
produce more understandable expressions need to approximate it, in
our case by expanding in $m/T$.  We need to be sure that terms
dropped at Matsubara frequencies do not alter the physics which
depends on the appropriate analytic continuation, not simply on
the values at Matsubara frequencies.

\subsubsection{The free case.}

To understand our approach, let us take the universal example of
an exactly solvable quantum field theory, a free scalar field.  We
can compare the analytic continuation of approximations to the ITF
version against the known exact result at real energies.
Of course, in this case it seems a
little stupid to make any approximation, such as DR, because we
know the answer exactly. However, it will  help us  to see how we
can be misled by DR.

The ITF propagator is given in \tref{acfreeit}. The unique
generalised free propagator, the one obeying the BM conditions
\tref{BMcond}, is just \tref{freerp2}.  Suppose now that we were
interested in the limit $\omega_p\ll 2\pi T$ where
$\omega_p^2=p^2+m^2$ (large distances and inverse masses compared
to $\beta$ or ``soft, long wavelength'' modes). This is what we
will mean by high $T$ here (for the full interacting theory we
will also take $T\gg m$, the fermion mass, see main text.). In
that limit we find an approximate form for the propagator
$\Delta_1$, much as one has in actual calculations. This is given
by
\begin{equation}
 \Delta_T^0 (i\omega_n,p) \approx
 \Delta^0_{T1}(i\omega_n,p)
 \left[1+\Od\left(\frac{\omega_p^2}{(2\pi T)^2}\right)\right]
  , \; \; \;
 \Delta^0_{T1}(i\omega_n,p) := \frac{\delta_{n0}}{\omega_p^2}
\label{itdr}
\end{equation}
i.e., this is the DR approximation and we pick up only the zero
mode ($n\neq 0$ represent ``heavy'' modes compared to $\omega_p$).
Now, if we find the unique generalised propagator, $\Delta_1$,
associated through \tref{BMcond} to the IT function $\Delta_{T1}$,
this analytic continuation of the DR propagator $\Delta_1$ is
\begin{equation}\Delta_1 (\omega,p) =
 0 \quad \mbox{for} \ \im \omega\neq 0,\pi \quad \Longrightarrow
 \quad
 \Delta_1^R (\omega,p) = \Delta_1^A (\omega,p)=
 0,\label{wrongac}
\end{equation}
since it is the behaviour of the IT function near the limit point
at $n = \pm \infty$ that controls the continuation, not the value
at any one isolated finite $n$ Matsubara energies.

This is clearly not a very good approximation for the real-energy
free propagator and is in fact inconsistent. First of all, we know
that the true answer for the retarded propagator  (\ref{freerp})
{\it is $T$-independent}. Thus, if we continue first and then take
the high $T$ limit, the answer is still (\ref{freerp}). Therefore,
the DR and analytic continuation operations do not commute in this
case. Second, even though (\ref{itdr}) is a good and well
controlled approximation to the IT propagator at high $T$, that is
not the case for its real-energy version in (\ref{wrongac}) when
compared to the true answer in (\ref{freerp}). For any arbitrarily
small $\omega$, they differ by $\omega_p^{-2}$, which is {\it not
small} at high $T$. On the other hand for $\omega>2\pi T$ the
difference between the two is $\omega_p^{-2}\times
\Od\left(\omega_p^2/(2\pi T)^2\right)$ which is negligible to
leading order in DR. Therefore, the conclusion is that by
performing this DR style continuation, up to the accuracy of our
approximations we are missing the right behaviour of the retarded
propagator  for frequencies $\omega\lesssim 2\pi T$.

Let us see if we can just solve the problem by making a better
approximation. Consider $\Delta_{T2}$ where
\begin{eqnarray}
 \Delta_T^0(i\omega_n,p) &=&
 \Delta_{T2}(i\omega_n,p)
\left[1+\Od\left(\frac{\omega_p^2}{(2\pi T)^2}\right)\right],
 \\
 \Delta_{T2}(i\omega_n,p) &=&
 \frac{\delta_{n0}}{\omega_p^2}
  + \sum_{k\neq 0} \frac{\delta_{nk}}{(2\pi k T)^2}  .
\label{itdr2}
\end{eqnarray}
This is an improvement over the simple DR form of \tref{itdr}. Now
{\it all the modes} appear as we take the leading high temperature
term for each mode rather than working to a fixed accuracy for all
terms. Notice however that the $n=0$ value does not fit the simple
functional form we have given for the $n \neq 0 $ heavy modes.
Thus we should not be surprised when the unique {\em analytic}
function obeying the BM conditions \tref{BMcond}, when trying to
match $\Delta_{T2}$, does not fit the $n=0$ term. The continuation
is controlled purely by the analytic behaviour of the Matsubara
sequence near the limit points of $\pm i \infty$, and here we find
\beq
 \Delta_2(\omega,p) =-\frac{1}{\omega^2}.
\label{wrongac2}
\end{equation}
The hard frequency behaviour is now a much better approximation to
that of the exact answer, but it is still a bad approximation for
the soft frequencies. This is not surprising as we have not used
the one soft frequency value we know.

The problem does not lie in the mathematical process of analytic
continuation. Both \tref{wrongac} and \tref{wrongac2} satisfy the
BM conditions \tref{BMcond} and they are therefore the unique
continuations to the relevant generalised propagators of
$\Delta_{T1}$ and $\Delta_{T2}$ respectively. On the other hand
the same is true for the exact answer (\ref{freerp}). Therefore,
we have three apparently different continuations of the high
temperature ITF propagator. This is not a contradiction, because
we are dealing with {\it approximate} expressions for the IT
propagator. Therefore, we are free to choose the continued version
which gives a better approximation also for real frequencies. The
problem is how can we specify the best continuation in a general
case where we do not have the exact analytic answer? Is there any
extra condition over and above those of Baym and Mermin
\tref{BMcond} which we can demand?

The answer is that we should demand that the continuation also
passes through the $n=0$ mode value, so incorporating the
information about the $n= 0$ mode. This value is unimportant for
the continuations, since it is the behaviour of the ITF function
near the limit points at $\pm i \infty$ which controls them.
However, the $n=0$ value is the only direct information we have on
the soft energy region, so we ought to use it if we expect to get
that physics correct. However, the continuation is fixed uniquely
by the value of the functions at hard Matsubara frequencies and so
we can not just choose its value at zero energy. What we must do
is exploit the fact that we are working with approximations and
alter the hard frequency values by small amounts equivalent to the
inherent errors of the approximation.  We will do this until we
have a set of values at Matsubara frequencies which give an
analytic continuation which agrees up to the errors of the
approximation scheme at all Matsubara frequencies with the known
values, including the zero energy one.

Let's see how this works for the free propagator. For that purpose
we recognise that adding a small energy independent constant will
do the trick, but there is no unique choice.  In this case one can
see from the functional form for the ITF propagator what the best
choice is.  Thus, adding `small' corrections in a high temperature
regime to the form $\Delta_2$, we find a third form
\begin{equation} \Delta_T^0(i\omega_n,p) =
 \Delta_3(i\omega_n,p)
\left[1+\Od\left(\frac{\omega_p^2}{(2\pi T)^2}\right)\right] ,
\;\;\;
 \Delta_3(i\omega_n,p) =
 \frac{\delta_{n0}}{\omega_p^2}
  + \sum_{k\neq 0} \frac{\delta_{nk}}{(2\pi k T)^2+\omega_p^2}  .
\label{itdr2b}
\end{equation}
This is not a unique choice and here we have exploited the
information in the ITF form.  We therefore add a new condition for
finding the unique generalised propagator through analytically
continuing an IT propagator, namely
\begin{equation}
\Delta(i\omega_n,p) \approx \Delta_T(i\omega_n,p) \; \; \; \forall
\; n \geq 0.
 \label{EGRScond}
\end{equation}
Note that this condition demands that $\Delta_R(0,p) =
\Delta_A(0,p)$.  However, this is guaranteed by the equal time
commutation relations for two-point functions.  An equivalent
identity is true for higher-point functions \cite{TSEzegf}.

In principle, all choices which differ by small amounts for each
Matsubara value should give equivalent physical results, again up
to the accuracy of the approximation scheme. However, as we have
seen in section \ref{sec:dr}, in addition to the mathematical
conditions we have just discussed, often one has to demand that
the physical answer is meaningful. That is the case for the
long-time behaviour of the conductivity discussed in section
\ref{phys}.


\section{Algebraic analysis of the sine-Gordon propagator}\label{sgpropalg}

Here we will give more details about some of the calculations
performed in section \ref{sec:sgpropit}.

First, we will show how to arrive to (\ref{sgitprop1}). Using
(\ref{sggf}) with the SG generating functional given in
 \cite{gs99} and taking the $\mu_0\rightarrow 0^+$ limit, we find

\begin{eqnarray}
 \Delta_T(x,\tau)
 &=&
 \Delta_T^0(x,\tau)
 +
 \frac{Z_0^B (T)}{Z_{SG} (T)}
 \sumno
 \Gamma_{2n} (X_1,  \ldots, X_{2n} )
 \\
 \Gamma_{2n} (X_1,  \ldots, X_{2n} )
 &=&-\left(\frac{\lambda}{4\pi}\right)^2
 \left(\frac{1}{n!}\right)^2
 \left[ \frac{\alpha}{2\lambda^2}
 \left(\frac{T} {\rho}\right)^{\lambda^2/4\pi} \right]^{2n}
 \left( \prodjintc \right),
 \nonumber\\
 &\times&
 \left[\sum_{l=1}^{2n}
 \ln \{ [ Q^2(x_l,\tau_l)]^{\epsilon_l \lambda /4\pi } \} \right]
 \left[\sum_{m=1}^{2n}
 \ln \{ [ Q^2(x_m-x,\tau_m-\tau)]^{\epsilon_m\lambda / 4\pi} \} \right]
 \nonumber\\
 && \times
 \nonumber \\ &&
 \left( \prod_{j=2}^{2n}\prod_{k=1}^{j-1} \left[
 Q^2(x_j-x_k,\tau_j-\tau_k)                
 \right]^{\epsilon_j\epsilon_k \lambda^2/4\pi} \right).
 \label{sgitprop1b2}
\end{eqnarray}

In order to obtain (\ref{sgitprop1}),  we note that in the
 $l$ and $m$ sums in the above integral we can always relabel
 variables so that $x_j$ becomes {\it any} of the $x_1,\dots,x_n$
 if $j<n$ and so on for $x_j$ with $n<j<2n$. The
  term $\prod_{1 \leq k < j \leq 2n}\left[
Q^2(x_j-x_k,\tau_j-\tau_k)\right]^{\epsilon_j\epsilon_k
\lambda^2/4\pi}$ remains invariant under such relabelling and also
under the exchange $x_j\leftrightarrow x_{j+n}$, $\forall
j=1,\dots,n$. In this way, we end up with (\ref{sgitprop1}), where
the $\ln Q^2$ factors are written only in terms of $x_1,x_2$ and
 $x_{2n}$.

Note that (\ref{sgitprop1b2}) is $\mu_0$ independent. For
convenience, in (\ref{sgitprop1}) we have written the final
expression in terms of $\Delta_0$ again. They are equivalent in
the $\mu_0\rightarrow 0^+$ limit. Let us focus now on the UV
behaviour of (\ref{sgitprop1b2}), i.e, small $x,\tau$. This
corresponds to the behaviour of the integrand in
 near the regions where $Q^{2}\rightarrow 0^{+}$
in denominators and logs (see a similar discussion in \cite{gs99}
for the partition function).  On the other hand,
$Q^2(x,\tau)\approx \pi^2 T^2 (x^2+\tau^2)$ for
$(x,\tau)\rightarrow (0,0)$. Thus, clearly the most divergent
contributions arise when $x=\tau=0$ and come from  integrals like:

\begin{equation}
\int_{T,y\sim 0} d^2y \int_{T,y'\sim y} d^2y'
\frac{1}{\left[Q^2(y-y')\right]^{\lambda^2/4\pi}} \ln Q^2 (y) \ln
Q^2 (y')\approx \int_{r\sim 0} dr \frac{\ln^2
r}{r^{[\lambda^2/2\pi -1}]},
\end{equation}
where $\int_T d^2 y\equiv\int_0^\beta dy_0
\int_{-\infty}^{+\infty} dy_1$ and $r^2=y_0^2+y_1^2$.
 Clearly, the above integral
 converges near $r=0$ for $\lambda^2<4\pi$.

Next, we will give more details on the derivation of
(\ref{itsefac}). When taking the Fourier transform of
(\ref{sgitprop1b2}) we realise that the only dependence with
$(\tau,x)$ appears in $\sum_{k=1}^{2n} \epsilon_k \ln
Q^2(x_k-x,\tau_k-\tau)=-4\pi\sum_{k=1}^{2n} \epsilon_k \Delta_T^0
(x_k-x,\tau_k-\tau)$, so that we can write (\ref{sgap1}) with

\begin{equation} \Sigma_T (i\omega_n,p)
 = \frac{\lambda^2}{4\pi}  \frac{Z_0^B (T)}{Z_{SG} (T)}
\sumno \left(\frac{1}{n!}\right)^2
\left[\frac{\alpha}{2\lambda^2}\left(\frac{T}
{\rho}\right)^{\lambda^2/4\pi}\right]^{2n} \Sigma_T^n
 (i\omega_n,p),
 \label{sgap2a}
\end{equation}
and \ba
 \Sigma_T^n (i\omega_m,p)\Delta_T^0(i\omega_m,p)
 =
 \prodjintc \prod_{1 \leq k < j \leq 2n}
 \left[
 Q^2(x_j-x_k,\tau_j-\tau_k)\right]^{\epsilon_j\epsilon_k
 \lambda^2/4\pi}
 \nonumber
 \\
 \times
 \left[\sum_{j=1}^{2n}
 \epsilon_j \ln Q^2(x_j,\tau_j)\right] \left[\sum_{k=1}^{2n}
 \epsilon_k e^{i\omega_m \tau_k}e^{-ipx_k}\right].
 \label{sgap2}
\end{eqnarray}

Let us now write the expression (\ref{sgap2}) in a slightly
different form,
 more convenient for our purposes. First, we relabel the variables
 in the $j$ and $k$ sums as we have just done with
 (\ref{sgitprop1b2}). Thus, separating the terms with $j=k$ and $j\neq k$ in the sum, we
can write: \ba
 \left[\Sigma_T^n \Delta_T^0\right] (i\omega_m,p)&=&
2\prodjintc \prod_{1 \leq k < j \leq 2n}\left[
Q^2(x_j-x_k,\tau_j-\tau_k)\right]^{\epsilon_j\epsilon_k
\lambda^2/4\pi}  \ln Q^2(x_1,\tau_1)\nonumber\\ &\times&\left\{n
e^{i\omega_m \tau_1}e^{-ipx_1}+n(n-1) e^{i\omega_m
\tau_2}e^{-ipx_2}-n^2 e^{i\omega_m \tau_{2n}}e^{-ipx_{2n}}.
\right\}\label{itseint}\end{eqnarray}

Now let us change variables to:
\begin{equation}x'_{2n}=x_{2n}-x_{2n-1}, \ \dots, \ x'_2=x_2-x_1, \
x'_1=x_1,
\label{cov}\end{equation} so that

\be x_j=\sum_{l=1}^j x'_l \quad , \quad x_j-x_k=\sum_{l=k+1}^j
 x'_l \quad (k<j<2n), \label{invcov}\end{equation} and so on for $\tau'_{2n},\dots
\tau'_1$. Note that with this change of variable,
\begin{equation} \prod_{1 \leq k < j \leq 2n}\left[
 Q^2(x_j-x_k,\tau_j-\tau_k)\right]^{\epsilon_j\epsilon_k
 \lambda^2/4\pi}=\prod_{1 \leq k < j \leq 2n}\left[ Q^2(\sum_{l=k+1}^j
 x'_l,\sum_{l=k+1}^j \tau'_l)\right]^{\epsilon_j\epsilon_k
 \lambda^2/4\pi}
\end{equation}
is independent of $(x'_1,\tau'_1)$. This is just a consequence of
two-dimensional translation invariance or, equivalently, total
energy-momentum conservation in any diagram, which becomes
manifest with this change of variables. Thus, we can {\it separate
the free propagator} $\Delta_T^0 (i\omega_m,p)$ in the r.h.s of
(\ref{itseint}) yielding (\ref{itsefac}).


\newpage
\typeout{--- No new page for bibliography ---}

\end{document}